\documentclass[useAMS,usenatbib]{mn2e}
\usepackage{graphics}
\usepackage{times}
\def\k{\mbox{\boldmath $k$}}
\def\r{\mbox{\boldmath $r$}}
\def\omm{\mbox{$\Omega_{\rm m}$}}
\def\omb{\mbox{$\Omega_{\rm b}$}}
\def\omk{\mbox{$\Omega_{\rm k}$}}
\def\oml{\mbox{$\Omega_\Lambda$}}
\newcommand{\hompc}{\,h\,{\rm Mpc}^{-1}}
\newcommand{\mpcoh}{\,h^{-1}\,{\rm Mpc}}

\begin{document}

\title[Measuring the BAO scale]{Measuring the Baryon Acoustic
  Oscillation scale using the SDSS and 2dFGRS}

\author[W.J. Percival et al.]{
\parbox{\textwidth}{
Will J. Percival$^{1}$\thanks{E-mail: will.percival@port.ac.uk}, 
Shaun Cole$^{2}$, 
Daniel J. Eisenstein$^{3}$,
Robert C. Nichol$^{1}$, 
John A. Peacock$^{4}$,
Adrian C.\ Pope$^{5}$,
Alexander S.\ Szalay$^{6}$}
\vspace*{4pt} \\
$^{1}$ Institute of Cosmology and Gravitation, University of
Portsmouth, Portsmouth, P01 2EG, UK \\
$^{2}$Department of Physics, University of Durham, Science
Laboratories, South Road, Durham DH1 3LE, UK \\
$^{3}$Steward Observatory, University of Arizona, 933
N. Cherry Ave., Tucson, AZ 85121, USA \\
$^{4}$SUPA; Institute for Astronomy, University of Edinburgh, Royal
Observatory, Edinburgh EH9 3HJ, UK \\
$^{5}$Institute for Astronomy, University of Hawaii,
2680 Woodlawn road, Honolulu, HI 96822, USA \\
$^{6}$Department of Physics and Astronomy, The Johns Hopkins
University, 3701 San Martin Drive, Baltimore, MD 21218, USA}

\date{\today} 

\maketitle

\begin{abstract}
  We introduce a method to constrain general cosmological models using
  Baryon Acoustic Oscillation (BAO) distance measurements from galaxy
  samples covering different redshift ranges, and apply this method to
  analyse samples drawn from the SDSS and 2dFGRS. BAO are detected in
  the clustering of the combined 2dFGRS and SDSS main galaxy samples,
  and measure the distance--redshift relation at $z=0.2$. BAO in the
  clustering of the SDSS luminous red galaxies measure the
  distance--redshift relation at $z=0.35$. The observed scale of the
  BAO calculated from these samples and from the combined sample are
  jointly analysed using estimates of the correlated errors, to
  constrain the form of the distance measure
  $D_V(z)\equiv[(1+z)^2D_A^2cz/H(z)]^{1/3}$. Here $D_A$ is the angular
  diameter distance, and $H(z)$ is the Hubble parameter. This gives
  $r_s/D_V(0.2)=0.1980\pm0.0058$ and $r_s/D_V(0.35)=0.1094\pm0.0033$
  (1$\sigma$ errors), with correlation coefficient of $0.39$, where
  $r_s$ is the comoving sound horizon scale at recombination. Matching
  the BAO to have the same measured scale at all redshifts then gives
  $D_V(0.35)/D_V(0.2)=1.812\pm0.060$. The recovered ratio is roughly
  consistent with that predicted by the higher redshift SNLS
  supernovae data for $\Lambda$CDM cosmologies, but does require
  slightly stronger cosmological acceleration at low redshift. If we
  force the cosmological model to be flat with constant $w$, then we
  find $\omm=0.249\pm0.018$ and $w=-1.004\pm0.089$ after combining
  with the SNLS data, and including the WMAP measurement of the
  apparent acoustic horizon angle in the CMB.
\end{abstract}

\begin{keywords}
  cosmology: observations, distance scale, large-scale structure of
  Universe
\end{keywords}

\section{Introduction}

The physics governing the production of Baryon Acoustic Oscillations
(BAO) in the matter power spectrum is well understood
\citep{silk68,peebles70,sunyaev70,bond84,bond87,holtzman89}. These
oscillatory features occur on relatively large scales, which are still
predominantly in the linear regime; it is therefore expected that
BAO should also be seen in the galaxy distribution
\citep{meiksin99,springel05,seo05,white05,eisenstein07}. Consequently,
BAO measured from galaxy surveys can be used as standard rulers to
measure the geometry of the Universe through the distance--redshift
relation \citep{blake03,seo03}.

BAO have now been convincingly detected at low redshift in the 2dFGRS
and SDSS galaxy samples \citep{cole05,eisenstein05,huetsi06}. With the
latest SDSS samples they are now detected with sufficient signal to
use BAO alone to measure cosmological parameters \citep{percival07a}.
This has emphasised the importance of accurate models for BAO in the
galaxy power spectrum. On small scales, BAO will be damped due to
non-linear structure formation \citep{eisenstein07}. Given the
accuracy of current data, uncertainty in the exact form of this
damping is not important, but it will become so for future data
sets. On larger scales, there is currently no theoretical reason to
expect systematic distortions greater than $\sim1\%$ in the BAO
positions between the galaxies and the linear matter distribution
\citep{seo03,springel05,seo07,angulo07}.  Claims of $>1\%$ changes in
the BAO position have used non-robust statistical measures of the BAO
scale, such as the position of the bump in the correlation function,
or peak locations in the power spectrum
\citep{smith07a,smith07b,crocce07}. These are easily affected by
smooth changes to the galaxy clustering amplitude as a function of
scale. In this paper, we use a more robust approach: the BAO scale is
defined via the locations where the BAO cross a smooth fit to the
power spectrum.

Ideally we would use the BAO within two galaxy redshift surveys
covering different narrow redshift slices to test a cosmological model
using the following procedure:
\begin{enumerate}
\item Convert from galaxy redshift to distance assuming the
  cosmological model to be tested.
\item Calculate the galaxy power spectra for the two samples.
\item Measure the oscillations in each power spectrum around the known
  smooth underlying power spectrum shape.
\item Test whether the change in scale between the two observed BAO
  positions agrees with that expected for this cosmological model.
\end{enumerate}
Unfortunately, a number of complications prevent such a simple
procedure from being used. In particular, this method requires a
distance--redshift relation to be specified prior to measuring the BAO
positions; but the errors and the effect of the survey selection
function depend on this assumption, and these are computationally
expensive to measure for many different models. In recent analyses
\citep{percival01,cole05,tegmark06}, a fiducial cosmological model has
been used to estimate the power spectrum, and the effect of this on
the recovered shape of the power has been tested. However, when
providing BAO distance scale measurements we need to allow for the
change in the distance--redshift relation. In this paper, we calculate
the power spectrum for a fiducial cosmology, and interpret these data
as if the model cosmology had been analysed (incorrectly) assuming the
fiducial model, therefore allowing for this effect. This procedure
gives better noise properties for the derived parameters than
recalculating the BAO for each model.

We test models against the data for general smooth forms of the
distance--redshift relation, parametrised by a small number of
nodes. This allows for surveys covering a range of redshifts, and has
the advantage of allowing derived constraints to be applied to any
model provided that it has such a smooth relation.  Our ``ideal''
method also required us to know the power spectrum shape so we could
extract the BAO. In this paper, we do not model this shape using
linear CDM models. To immunise against effects such as scale-dependent
bias, non-linear evolution, or extra physics such as massive
neutrinos, we instead model the power spectrum shape by fitting with a
cubic spline.

The method is demonstrated by analysing galaxy samples drawn from the
combined SDSS and 2dFGRS (Section \ref{sec:analysis}). Results are
presented in Sections \ref{sec:results} \&~\ref{sec:cosmo}, and
discussed in Section~\ref{sec:discussion}. This application is novel,
as we combine the 2dFGRS and SDSS galaxy samples before calculating
power spectra (the two data sets are introduced in
Section~\ref{sec:data}). The blue selection in the 2dFGRS and the red
selection in the SDSS galaxies emphasise 
different classes of galaxies with different large-scale biases -- but these
can be matched using a relative bias model leading to the same
large-scale power spectrum amplitudes
\citep{cole05,tegmark06,percival07b}. If there is scale-dependent
bias, then the shape of the power spectrum calculated from the
combined sample will be an average of the two individual power
spectra, because we are selecting a mix of galaxy pairs. The exact
mix will change with scales, but, this is not
expected to be a significant concern for the BAO positions in the
power spectra; these should be the same across all data sets, although
there will be an effect on the damping of BAO on small scales (this is discussed
in Section~\ref{sec:bao_model}).

\section{The Data}  \label{sec:data}

\subsection{The SDSS data}

The public SDSS samples used in this analysis are the same as
described in \citet{percival07b}. The SDSS
\citep{york00,adelman06,blanton03,fukugita96,gunn98,gunn06,hogg01,ivezic04,pier03,smith02,stoughton02,tucker06}
Data Release 5 (DR5) galaxy sample is split into two subsamples: there
are 465789 main galaxies \citep{strauss02} selected to a limiting
extinction-corrected magnitude $r<17.77$, or $r<17.5$ in a small
subset of the early data from the survey. In addition, we have a
sample of 56491 Luminous Red Galaxies (LRGs; \citealt{eisenstein01}),
which form an extension to the survey to higher redshifts
$0.3<z<0.5$. Of the main galaxies, 21310 are also classified as LRGs,
so our sample includes 77801 LRGs in total. Although the main galaxy
sample contains significantly more galaxies than the LRG sample, the
LRG sample covers more volume. The redshift distributions of these two
samples are fitted as described in \citet{percival07b}, and the
angular mask is determined using a routine based on a {\sc HEALPIX}
\citep{gorski05} equal-area pixelization of the sphere
\citep{percival07b}. In order to increase the volume covered at low
redshift, we include the 2dFGRS sample, which for simplicity has been
cut to exclude angular regions covered by the SDSS samples.

\subsection{The 2dFGRS data}

The full 2dF Galaxy Redshift Survey (2dFGRS) catalogue contains
reliable redshifts for 221\,414 galaxies selected to an
extinction-corrected magnitude limit of approximately $b_J=19.45$
\citep{colless01,colless03}. For our analysis, we only select regions
not covered by the SDSS survey, and we do not include the random
fields, a set of 99 random 2~degree fields spread over the full
southern galactic cap. This leaves 143\,368 galaxies in total. The
redshift distribution of the sample is analysed as in \citet{cole05},
and we use the same synthetic catalogues to model the unclustered
expected galaxy distribution within the reduced sample.

The average weighted galaxy densities in the SDSS and 2dFGRS
catalogues were calculated separately, and the overall normalisation
of the synthetic catalogues were matched to each catalogue separately
using these numbers (see, for example, \citealt{cole05} for
details). The relative bias model described in \citet{percival07b} was
applied to the SDSS galaxies and the bias model of \citet{cole05} was
applied to the 2dFGRS galaxies. These normalise the large-scale
fluctuations to the amplitude of $L_*$ galaxies, where $L_*$ is
calculated separately for each survey. We therefore include an extra
normalisation factor to the 2dFGRS galaxy bias model to correct the
relative bias of $L_*$ galaxies in the different surveys. This was
calculated by matching the normalisation of the 2dFGRS and SDSS bias
corrected power spectra for $k<0.1\hompc$. 2dFGRS galaxies at a single
location were all given the same expected bias, rather than having
biases matched to their individual luminosities. This matches the
method used for the SDSS, and makes the calculation of mock catalogues
easier.

\section{BAO in the Galaxy Power Spectrum}  
\label{sec:bao_model}

In this section, we consider the relation between BAO measured from
the galaxy distribution, and BAO in the linear matter distribution. We
define the linear BAO as
\begin{equation}
  B_{\rm lin}(k) \equiv \frac{[T_{\rm full}(k)]^2}{[T_{\rm no\ osc}(k)]^2},
  \label{eq:Blin}
\end{equation}
which oscillates around $B_{\rm lin}(k)=1$. $T_{\rm full}(k)$ is the
full linear transfer function, while $T_{\rm no\ osc}(k)$ is the same
without the sinusoidal term arising from sound waves in the early
universe. In the fitting formulae provided by \citet{eisenstein98},
this term is given by their equation 13, a modified sinc
function. Note that $T_{\rm no\ osc}(k)$ contains the change in the
overall shape of the power spectrum due to baryons affecting the small
scale damping of perturbations, just not the oscillatory
features. $B_{\rm lin}(k)$ can be considered as a multiplicative
factor that corrects the smooth power spectrum to provide a full
model.

Within the halo model \citep{seljak00,peacock00,cooray02}, the
real-space galaxy power spectrum is related to the linear power
spectrum by the addition of an extra smooth term, and multiplication
by a smooth, possibly scale dependent, galaxy bias $b(k)$
\begin{equation}
  P_{\rm obs}(k) = b^2(k)P(k)_{\rm lin} + P(k)_{\rm extra}.
  \label{eq:pk_obs}
\end{equation}
The $b^2(k)$ term can also be thought of as equivalent to the Q-model
of \citet{cole05}, used to model the transition between the linear
matter power spectrum and observed galaxy power spectra. The form of
Equation (\ref{eq:pk_obs}) matches that calculated by
\cite{scherrer98} from a general hierarchical clustering
argument. $b^2(k)$ and $P(k)_{\rm extra}$ are required to be slowly
varying functions of $k$ such that we can extract the BAO signal as
follows. Substituting Equation (\ref{eq:Blin}) into Equation
(\ref{eq:pk_obs}), and writing $P_{\rm lin}(k)=Ak^n[T_{\rm
  full}(k)]^2$ gives
\begin{equation}
  P_{\rm obs}(k) = Ab^2(k)k^n B_{\rm lin}(k) [T_{\rm no\ osc}(k)]^2 
    + P(k)_{\rm extra}.
\end{equation}
We extract BAO from this observed power spectrum by dividing by a
smooth model that, without loss of generality, we can choose to be
\begin{equation}
  P(k)_{\rm smooth} = Ab^2(k)k^n [T_{\rm no\ osc}(k)]^2 + P(k)_{\rm extra},
  \label{eq:pk_smooth}
\end{equation}
so the oscillations in $P_{\rm obs}(k)/P(k)_{\rm smooth}$ are
\begin{equation}
  B_{\rm obs}(k) = g(k) B_{\rm lin}(k) + [1-g(k)],
  \label{eq:Bobs}
\end{equation}
where
\begin{equation}
  g(k) = \frac{Ab^2(k)k^n [T_{\rm no\ osc}(k)]^2}
  {Ab^2(k)k^n [T_{\rm no\ osc}(k)]^2 + P(k)_{\rm extra}}
\end{equation}
is smooth. The $k$-scales where $B_{\rm obs}(k)=1$ occur where $B_{\rm
  lin}(k)=1$, showing that the oscillation wavelength is unchanged by
the translation given by Equation (\ref{eq:pk_obs}). However, the
positions of the maxima and minima will change as $g(k)$ is expected
to be asymmetric around the extrema. In fact, the detailed shape and
amplitude of this damping term will depend on the cosmological model
and on the properties of the galaxies being
analysed. \citet{eisenstein07} have shown that $g(k)$ can be
approximated as a Gaussian convolution in position-space with
$\sigma_g=10\mpcoh$ for low redshift galaxies.  For our default
results presented in this paper, we fix the damping model to be
Gaussian with $\sigma_g=10\mpcoh$, which is assumed not to change
significantly over the redshifts or galaxy types used in the
analysis. We consider variations in the BAO damping model in Section
\ref{sec:bao_test}. Equation (\ref{eq:Bobs}) shows that the observed
power spectrum is constructed from a smooth component (Equation
\ref{eq:pk_smooth}), and a multiplicative damped BAO model (Equation
\ref{eq:Bobs}). We assume that such a decomposition can be performed
for power spectra measured from galaxy samples drawn from the 2dFGRS
and SDSS.

We model $P(k)_{\rm smooth}$ as a 9 node cubic spline \citep{press92}
designed to be able to match the overall shape of the power spectrum
(i.e. to fit Equation \ref{eq:pk_smooth}), but not the BAO. The 9
nodes were fixed empirically at $k=0.001$, and $0.025\le k\le0.375$
with $\Delta k=0.05$. A cubic spline $\times$ BAO model with this node
separation was found to be able to fit model linear power spectra by
\citet{percival07a} and can match all of the power spectra presented
in this paper without leaving significant residuals. The $\chi^2$
values of the fits are all within the expected range of values. We
also consider an offset node distribution in Section
\ref{sec:bao_test}. The spline curve can be taken as the definition of
``smooth'': only effects that cannot be modelled by such a curve will
affect the BAO positions. When fitting the observed BAO, we do not
attempt to extract the BAO and then fit different models to these
data, because the method by which the BAO are extracted might bias the
result. Instead we fit combined cubic spline $\times$ BAO models to
the power spectra, allowing the spline fit to vary with each BAO model
tested (this follows the method of \citealt{percival07a}).

We now consider how to model the BAO. \citet{blake03} suggest
modelling $B_{\rm lin}(k)$ using a simple damped sinusoidal two-parameter
function
\begin{equation}
  B_{\rm lin}(k) = 1+Ak
  \exp\left[-\left(\frac{k}{0.1\hompc}\right)^{1.4}\right]
  \sin\left(\frac{2\pi k}{k_A}\right),
  \label{eq:bao_bg03}
\end{equation}
where $k_A=2\pi/r_s$, and $r_s$ is the co-moving sound horizon scale at
recombination at scale factor $a_*$
\begin{equation}
  r_s =  \frac{1}{H_0 \omm^{1/2}} 
    \int_0^{a_*} \frac{c_S}{(a + a_{\rm eq})^{1/2}} \, da.
  \label{eq:rs} 
\end{equation}
Here, the amplitude $A$ is treated as a free parameter. In this paper,
we consider units $\mpcoh$, so working in these units $H_0\equiv100$
in Equation (\ref{eq:rs}). This simple function ignores issues such as
the propagation of the acoustic waves after recombination. Although
the sound speed drops radically at recombination, acoustic waves still
propagate until the end of the 'drag-epoch'. This leads to the
slightly larger sound horizon as measured from the low-z galaxy
clustering data than the CMB. To include such effects, we use a BAO
model extracted from a power spectrum calculated using the numerical
Boltzmann code {\sc CAMB} \citep{lewis00}, by fitting with a cubic
spline $\times$ BAO model.  For simplicity, we index our results based
on the sound horizon at recombination, $r_s$. In principle, there
could be small errors here (i.e. the large-scale structure to CMB
sound horizon ratio could be a function of cosmology), but the
combination of the current results and WMAP data mean that we are not
looking over that big a range of cosmological parameters. To test
this, we have applied the spline $\times$ BAO fit to {\sc CAMB} power
spectra for flat $\Lambda$CDM models with recombination sound horizon
scales covering the 2-$\sigma$ range of our best fit numbers
($\pm6\%$). We find that the input sound horizon at recombination is
recovered with less than 1\% error from these fits, showing that this
approximation is not important to current measurement precision.

For our default results, we extract the BAO model from a power
spectrum calculated assuming $\omm=0.25$, $\omb h^2=0.0223$ and
$h=0.72$. For these parameters $r_s=111.426\mpcoh$, calculated using
formulae presented in \citet{eisenstein98}. Small differences of
convention in computing the sound horizon scale can be accommodated by
simply scaling to match this value for these cosmological parameters.
If recovered bounds on $r_s$ are to be used to constrain models where
$r_s$ is not calculated using the formulae presented in
\citet{eisenstein98}, then our results should be shifted using the
difference between $r_s=111.426\mpcoh$ and the model recombination
sound horizon scale for $\omm=0.25$, $\omb h^2=0.0223$ and $h=0.72$.

\section{Observing The BAO Scale} \label{sec:obs_bao_scale}

\subsection{Narrow redshift shell surveys}

Suppose that a survey samples a narrow redshift shell of width $\Delta
z$ at redshift $z$. Furthermore, suppose that we are only interested
in the clustering of galaxies pairs with small separations. For a
given pair of galaxies, $\Delta z$ and the angular separation $\theta$
are fixed by observation, and we wish to measure the comoving
separation for different cosmological models.  In the radial
direction, separations in comoving space scale with changes in the
cosmological model as $dr_c/dz\simeq\Delta r_c / \Delta z = c/H(z)$,
where $r_c(z)\equiv\int c(1+z)\,dt$ is the comoving distance to a
redshift $z$. In the angular direction, the comoving galaxy separation
scales as $\Delta r_c = \Delta\theta (1+z)D_A$, where $D_A$ is the
standard angular diameter distance. Writing $S_k\equiv(1+z)D_A$,
\begin{equation}
S_k(z) =  \frac{c}{H_0}
  \cases{ 
    |\omk|^{-1/2}{\rm sinh}[\sqrt{\omk}H_0\,r_{\rm c}(z)/c] &if $(\omk>0)$, \cr
    H_0\,r_{\rm c}(z)/c &if $(\omk=0)$, \cr
    |\omk|^{-1/2}\sin[\sqrt{-\omk}H_0\,r_{\rm c}(z)/c] &if $(\omk<0)$. \cr
  }
  \label{eq:Sk}
\end{equation}
where $\omk=1-\Omega_0$ and $\Omega_0$ is the ratio of total to
critical density today. If we assume that the pairs of galaxies are
statistically isotropic, then we can combine the changes in scale and,
to leading order, the measured galaxy separations scale with the
cosmological model through the distance measure
$D_V(z)=[(1+z)^2D_A^2cz/H(z)]^{1/3}$. Here, we have introduced a
further factor of $z$ to match the definition of $D_V$ by
\citet{eisenstein05}: including functions of redshift does not change
the dependence of $D_V$ on different cosmological models. The position
of features in the real space 2-pt functions, the (dimensionless)
power spectrum and correlation function will approximately scale with
this distance measure. It is worth emphasising that this is only an
approximation, and would additionally be affected by redshift-space
distortions and other anisotropic effects.

Following these approximations, for a survey covering a narrow
redshift slice, the power spectrum $P(k)$ only needs to be calculated
for a single distance--redshift model. This is easiest if we assume a
flat cosmological model so we can set up a comoving Euclidean grid of
galaxies where BAO have the same expected scale in radial and angular
directions. The power spectrum for other models can be recovered by
simply rescaling the measured power in $1/D_V(z)$. Note that we could
have instead worked in dimensionless units $x/D_V(z)$, where the power
spectrum is independent of the comoving distance--redshift
relation. The position of the BAO in the power spectrum constrain $r_s
/ D_V(z)$, which is analogous to the peak locations in the Cosmic
Microwave Background (CMB) measuring $r_s/S_k(z_{\rm ls})$ (ignoring
the astrophysical dependencies of the peak phases), where $z_{\rm ls}$
is the redshift of the last scattering surface.

\subsection{Surveys covering a range of redshift}  \label{sec:survey_zrange}

We now consider what it means to measure the BAO scale in surveys
covering a range of redshifts. In this situation, the comoving
distance--redshift model assumed in measuring $\xi$ or $P(k)$ becomes
increasingly important. We first consider a simple survey covering two
redshift shells, and then extrapolate to more general surveys.

Consider measuring the correlation function as an excess of galaxy
pairs in a survey covering two redshift shells at redshifts $z_1$ and
$z_2$. Our estimate of the correlation function from the combined
sample will be the average of the correlation functions measured in
the two redshift bins, weighted by the expected total number of pairs
in each bin $W(z_i)$, and stretched by the distance $D_V(z_i)$. BAO in
the power spectrum correspond to a ``bump'' in the correlation
function, and the position of the bump scales with the BAO position,
and therefore measures $r_s / D_V(z_i)$. For two redshift slices, the
position of the bump in the combined correlation function depends on
the average position of the bumps in the correlations functions for
each slice, weighted by the total number of pairs in each bin. If
$D_V(z_1)$ is varied, then the same final BAO scale can be obtained
from the combined data provided that $D_V(z_2)$ is chosen such that
$[W(z_1)D_V(z_1)+W(z_2)D_V(z_2)]$ remains constant. Extending this
analysis to a large number of redshift shells, we see that the
measured BAO scale, assuming that this is measured from the mean
position of the bump in the correlation function, depends on
$r_s/\hat{D}_V$ where
\begin{equation}
  \hat{D}_V \equiv \int W(z) D_V(z)\,dz
  \label{eq:dvhat}
\end{equation}
Here, we do not have to worry about pairs of galaxies where the
galaxies are in different shells because of the small separation
assumption. The contributions from different redshifts $W(z)$, are
calculated from the weighted galaxy redshift distribution
squared. Because the weights applied to galaxies when analysing
surveys tend to upweight low density regions the BAO will, in general,
depend on a wider range of redshift than given by the radial
distribution of galaxies.

Now suppose that an incorrect comoving distance--redshift model
$\bar{D}_V(z)$ was assumed in the measurement of $\xi$ or
$P(k)$. Furthermore, suppose that this model $\bar{D}_V(z)$ has a
different shape to the true $D_V(z)$ but the same value of
$\hat{D}_V$. In this situation, our measurement of $r_s / \hat{D}_V$ is
unbiased with respect to the true value. What has changed is that the
BAO signal has been washed out: the recovered BAO in the power
spectrum are of lower amplitude, and the peak in the correlation
function broadens, because the BAO scales measured at different
redshifts are not in phase, although they sum so that their average
has the correct wave-scale. Note that if $\bar{D}_V$ matches the true
cosmological model, then there is no distortion of the BAO positions.

\subsection{Fitting the distance--redshift relation}

There are many different ways of parametrizing the distance--redshift
relation. For example, we could consider a cubic spline fit to
$r_c(z)$, $dr_c/dz$ or $D_V(z)$. For $\Lambda$ cosmologies the
comoving distance varies smoothly with redshift, and $D_V(z)$,
$r_c(z)$ and $dr_c/dz$ can all be accurately fitted by a spline with a
small number of nodes. In this paper, we fit $D_V(z)$ because of its
physical meaning in a simplified survey analysis on small scales;
but for non-flat cosmologies we cannot uniquely recover $r_c(z)$
from $D_V(z)$. This is not a problem because we only expect to measure
$D_V(z)$, and mocks calculated assuming the same $D_V(z)$, but with
different geometries, should give the same cosmological
constraints. Consequently, without loss of generality, we can assume
flatness when calculating the comoving distances from $D_V(z)$ in
order to create mock catalogues, and use
\begin{equation}
  r_c(z)^{\rm flat}=\left[3\int_0^z\frac{D_V^3(z')}{z'}\,dz'\right]^{1/3}.
\end{equation}

We now consider some of the practicalities of fitting the
distance--redshift relation.  Scaling of the distance--redshift
relation can be mimicked by ``stretching'' the measured power spectra
in $k$. Consequently, if we parametrize the distance--redshift model
by $N$ numbers, then power spectra only actually need to be calculated
for a set of $N-1$ values. For example, if the distance--redshift
model was parametrised by three nodes $D_V(z_1)$, $D_V(z_2)$ \&
$D_V(z_3)$, power spectra only need to be calculated for different
$D_V(z_2)/D_V(z_1)$ and $D_V(z_3)/D_V(z_1)$ values. Working in units
of $\mpcoh$ and fitting $dr_c/dz$, is one way of including such a
dilation of scale in the analysis: in these units the node at $z=0$ is
fixed $dr_c/dz|_{z=0}=c/H_0$, and only $N-1$ nodes are free to
vary. Allowing such a dilation at $z=0$, may not be the optimal choice
for the analysis of a survey at higher redshift.

\begin{figure}
  \centering
  \resizebox{0.9\columnwidth}{!}{\includegraphics{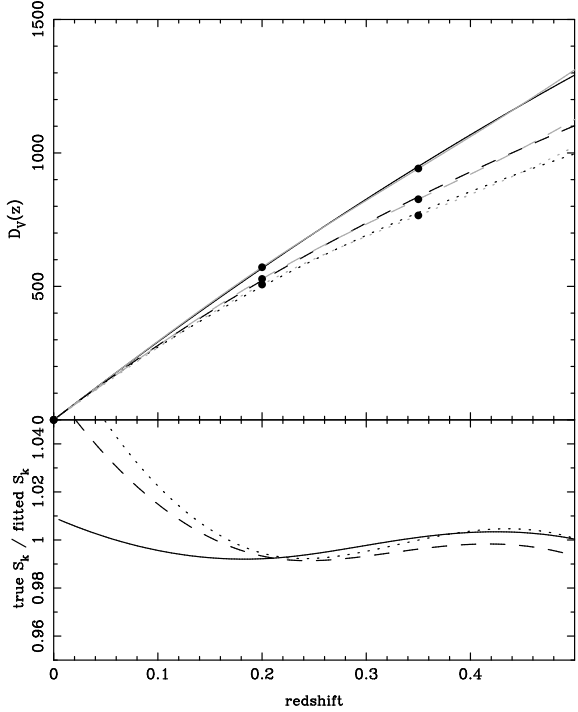}}
  \caption{The result of fitting to $D_V(z)$ using a cubic spline fit
    with three nodes at $z=0.0,0.2,0.35$ for $0<z<0.5$. We plot
    results for three cosmological models : $\Lambda$CDM ($\omm=0.25$,
    $\oml=0.75$, solid lines), SCDM ($\omm=1$, $\oml=0$, dotted
    lines), and OCDM ($\omm=0.3$, $\oml=0$, dashed lines). The upper
    panel shows the true values of $D_V(z)$ (black lines) compared
    with the spline fits (grey lines) with nodes (solid circles). The
    lower panel shows the resulting errors on $S_k$ as given by
    Equation (\ref{eq:Sk}). For the redshift range $z>0.15$, the error
    is $<1\%$. \label{fig:dp_2parfit}}
\end{figure}

By fitting the comoving distance (or a function of it), we hope to
provide measurements that can be easily applied to any set of
cosmological models, although we only have to analyse a small number
of comoving distance--redshift relations. The cosmological models that
can be tested require that the distance measure adopted can be well
matched by the parametrisation used. In this paper, we model $D_V(z)$
by a cubic spline fit with 2 nodes at $z=0.2$ and $z=0.35$:
consequently the results should only be used to delineate between
cosmological models where $D_V(z)$ is well modelled by such a
fit. Fig.~\ref{fig:dp_2parfit} shows fits of this form matched to a
selection of standard cosmological models (assuming a constant
weighted galaxy distribution with redshift). The error induced on the
comoving distance as a result of fitting $D_V(z)$ is small for these
models. The boundary conditions of the cubic spline are set so that
the second derivatives are zero at $z=0$ and $z=0.35$.

\subsection{Differential distance measurements}

In order to break the degeneracy between distance measurements at
different redshifts inherent in a single measurement of the BAO scale,
we need to analyse the BAO position in multiple power spectra or
correlation functions. This is true even if we are not in the regime
where the small separation assumption holds, although the degeneracy
would then be a more complicated function of the comoving distance
than $\hat{D}_V$ (Equation \ref{eq:dvhat}).

For the analysis of the 2dFGRS and SDSS DR5 galaxies presented in this
paper, the sample is naturally split into main galaxies (2dFGRS and
SDSS), SDSS LRGs, and the combination of the three samples. These
samples obviously overlap in volume, so the derived power spectra will
be correlated. However, using overlapping samples retains more
information than contiguous samples which would remove pairs across
sample boundaries. There is no double counting as each power spectrum
contains new information, and correlations between different power
spectra are included in the calculation of model likelihoods.

\subsection{Basic method} \label{sec:basic_method}

For each distance--redshift relation to be tested using the observed
BAO locations, we could recalculate the power spectrum and measure the
BAO positions. However, the likelihood of each model would not vary
smoothly between different models because the shot noise term in each
band-power varies in a complicated way with the distance--redshift
relation. This would give a ``noisy'', although unbiased, likelihood
surface.

An alternative approach is to fix the distance--redshift relation used
to calculate the power spectra. If this is different from the model to
be tested, the difference can be accounted for by adjusting the window
function - each measured data value has a different interpretation for
each model tested. One advantage of such an approach is that the shot
noise component of the data does not change with the model tested,
leading to a smoother and easier to interpret likelihood surface. The
primary difficulty is that the calculation of the window for each
model is computationally intensive. We now consider the mathematics
behind this approach.

Following \citet{FKP}, we define the weighted galaxy fluctuation field as
\begin{equation}
  f(\r)\equiv\frac{1}{N}w(\r)\left[n_g(\r)-\alpha n_s(\r)\right],
    \label{eq:Fr}
\end{equation}
where $n_g(\r)=\sum_j\delta(\r-\r_j)$ with $\r_j$ being the location
of the $j$th galaxy, and $n_s(\r)$ is defined similarly for the
synthetic catalogue with no clustering. Here $\alpha$ is a constant
that matches the average densities of the two catalogues (see, for example,
\citealt{PVP}), and $N$ is a normalization constant defined by
\begin{equation}
  N=\left\{\int d^3r\left[\bar{n}(\r)w(\r)\right]^2\right\}^{1/2}.
  \label{eq:N}
\end{equation}
$\bar{n}(\r)$ is the mean galaxy density, and $w(\r)$ is the weight
applied. The power spectrum of the weighted overdensity field $f(\r)$
is given by
\begin{equation}
  \langle |F(\k)|^2\rangle =
    \int d^3r\int d^3r' \langle f(\r)f(\r') \rangle e^{i\k\cdot(\r-\r')}.
  \label{eq:Fksq}
\end{equation}

The important term when substituting Equation (\ref{eq:Fr}) into
Equation (\ref{eq:Fksq}) is the expected 2-point galaxy density given
by
\begin{equation}
  \langle n_g(\r)n_g(\r') \rangle 
    = \bar{n}(\r)\bar{n}(\r')\left[1+\xi(\hat{\r}-\hat{\r'})\right]
    + \bar{n}(r)\delta_D(\r-\r').
  \label{eq:n_gn_g}
\end{equation}
If we analyse the galaxies using a different cosmological model to the
``true'' model, the 2-pt galaxy density depends on $\hat{\r}$ and
$\hat{\r'}$, the positions in the true cosmological model that are
mapped to positions $\r$ and $\r'$ when the survey is
analysed. Translating from the correlation function $\xi(\hat{r})$ to
the power spectrum $P(\hat{k})$ in the true cosmological model gives
\begin{equation}
  \xi(\hat{\r}-\hat{\r'})
    =\frac{1}{2\pi^2}\int P(\hat{\k})e^{-i\k.(\hat{\r}-\hat{\r'})}d^3\hat{k},
  \label{eq:xi_pk}
\end{equation}
which can be substituted into Equation (\ref{eq:n_gn_g}). Combining
Equations (\ref{eq:Fr} -- \ref{eq:xi_pk}) shows that the recovered
power spectrum is a triple integral over the true power. If
$\hat{\r}=\r$, this reduces to a convolution of the power spectrum
with a ``window function'' \citep{FKP}. If we now consider a piecewise
continuous true power spectrum
$P(k)=\sum_iP_i[\Theta(k)-\Theta(k-k_i)]$, where $\Theta(k)$ is the
Heaviside function, then the triple integral can be written as a
linear sum over $P_i$, $\langle|F(\k)|^2\rangle=\sum_iW_iP_i$. Because
the radial interpretation changes between actual and measured
clustering, spherically averaging the recovered power is no longer
equivalent to convolving the power with the spherical average of the
window function. Consequently, the window has to be estimated
empirically from mock catalogues created with different true power
spectra and analysed using a different cosmological model. The
empirical window function can be calculated including both the change
in cosmological model and the survey geometry.

\section{Analysis Of The SDSS and 2\lowercase{d}FGRS} 
  \label{sec:analysis}

\subsection{The observed BAO}  \label{sec:obs_bao}

\begin{figure}
  \centering
  \resizebox{0.9\columnwidth}{!}{\includegraphics{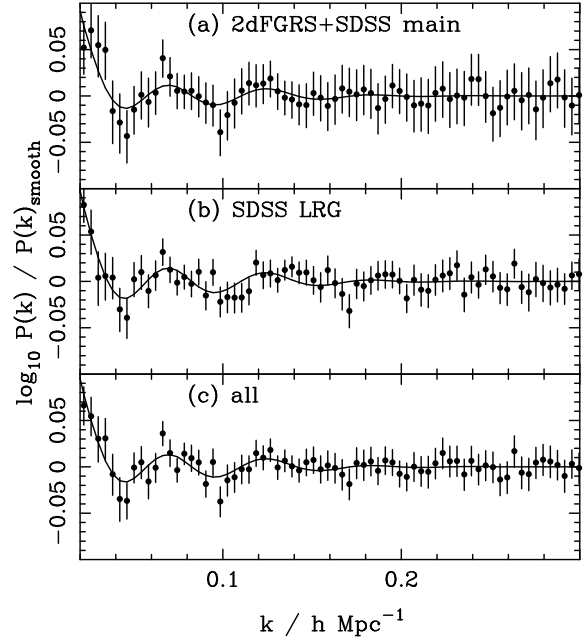}}
  \caption{BAO in power spectra calculated from (a) the combined SDSS
    and 2dFGRS main galaxies, (b) the SDSS DR5 LRG sample, and (c) the
    combination of these two samples (solid symbols with $1\sigma$
    errors). The data are correlated and the errors are calculated
    from the diagonal terms in the covariance matrix. A Standard
    $\Lambda$CDM distance--redshift relation was assumed to calculate
    the power spectra with $\omm=0.25$, $\oml=0.75$. The power spectra
    were then fitted with a cubic spline $\times$ BAO model, assuming
    our fiducial BAO model calculated using {\sc CAMB}, as described
    in Section (\ref{sec:bao_model}). The BAO component of the fit is
    shown by the solid line in each panel. \label{fig:bao_lrg_main}}
\end{figure}
Fig.~\ref{fig:bao_lrg_main} shows the BAO determined from power
spectra calculated for the combined sample of SDSS main galaxies and
2dFGRS galaxies, the SDSS LRG sample, and the combination of these
samples. The power spectra were calculated for $N=70$ band powers
equally spaced in $0.02<k<0.3\hompc$ using the method described in
\citet{percival07a}, assuming a flat $\Lambda$ cosmology with
$\omm=0.25$. Errors on these data were calculated from 2000 Log-Normal
(LN) density fields \citep{coles91} covering the combined volume, from
which overlapping mock samples were drawn with number density matched
to each galaxy catalogue. The distribution of recovered power spectra
includes the effects of cosmic variance and the LN distribution has
been shown to be a good match to the counts in cells on the scales of
interest $>10\mpcoh$ \citep{wild05}, so these catalogues should also
match the shot noise of the data. The catalogues do not include higher
order correlations at the correct amplitude for non-linear structure
formation, which are not included in the Log-Normal model. However,
the BAO signal comes predominantly from large-scales that are expected
to be in the linear or quasi-linear regimes, so these effects should
be small. Each catalogue was calculated on a $(512)^3$ grid covering a
$(4000\mpcoh)^3$ cubic volume. The recovered power spectra from these
mock catalogues were fitted with cubic spline $\times$ BAO fits as
described in Section~\ref{sec:bao_model}, and the errors on the BAO
were calculated after dividing by the smooth component of these fits.

We have fitted cubic spline $\times$ BAO models to the SDSS and 2dFGRS
power spectra using the method of \citet{percival07a}. For each
catalogue we have calculated the window function of the survey
assuming a flat $\Lambda$ cosmology with $\omm=0.25$ (using the method
described in \citealt{percival07a}), and the covariance matrix from
the LN catalogues, assuming that the power spectra band powers are
distributed as a multi-variate Gaussian. The power spectrum for each
sample was then fitted using cubic spline including or excluding the
multiplicative BAO model calculated using {\sc CAMB} as described in
Section \ref{sec:bao_model} for a flat $\Lambda$ cosmology with
$\omm=0.25$, $\omb h^2=0.0223$ \& $h=0.72$. All three samples are
significantly better fit by the models including BAO. For the combined
data, $-2\Delta\ln{\cal L}=9.6$, for the LRGs $-2\Delta\ln{\cal
  L}=7.4$, and for the SDSS main + 2dFGRS galaxies $-2\Delta\ln{\cal
  L}=5.9$ for the likelihood ratios between best-fit model power
spectra with BAO and without BAO.

\begin{figure}
  \centering
  \resizebox{0.9\columnwidth}{!}{\includegraphics{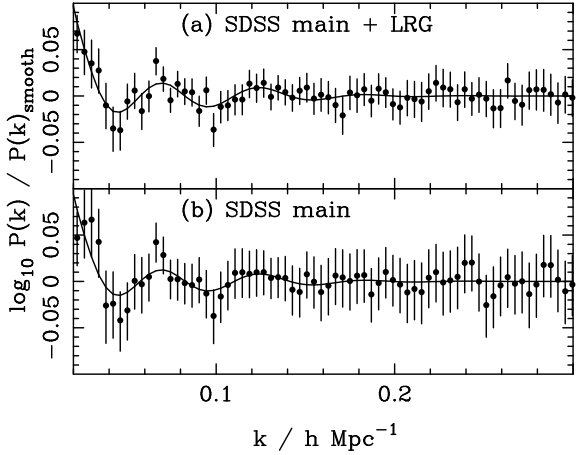}}
  \caption{As Fig.~\ref{fig:bao_lrg_main}, but for power spectra
    calculated from (a) the combined SDSS DR5 LRG and main galaxy
    sample, (b) the SDSS main galaxy sample. \label{fig:bao_sdss}}
\end{figure}
Including the 2dFGRS data reduces the error on the derived
cosmological parameters by approximately 25\% for our combined
analysis of three power spectra. The BAO calculated from just the SDSS
main galaxies and the combination of the SDSS main galaxies and the
LRGs are shown in Fig.~\ref{fig:bao_sdss}. From just the SDSS main
galaxies, $-2\Delta\ln{\cal L}=4.5$ for the likelihood ratios between
best-fit model power spectra with BAO and without BAO. There is no
change in the significance of the BAO detection from the combined SDSS
LRG and main galaxy sample from including the 2dFGRS galaxies.

The power spectra plotted in Fig.~\ref{fig:bao_lrg_main} are clearly
not independent. Some of the deviations between model and data in the
combined catalogue can be traced back to similar distortions in either
the main galaxy or LRG power spectra. The LRGs have a greater weight
when measuring the clustering of the combined sample on large-scales
compared with the lower redshift galaxies, while the low redshift
galaxies have a stronger weight when measuring the clustering on
smaller scales. The combined sample includes additional galaxy pairs
where the galaxies lie in different subsamples. All three samples also
cover different redshift ranges. As discussed in
Section~\ref{sec:obs_bao_scale} this means that they all contain
unique cosmological information. By simultaneously analysing all three
power spectra, allowing for the fact that they may be correlated, we
can therefore extract more cosmological information than by analysing
a single power spectrum.

\subsection{Fitting the distance--redshift relation}

\begin{figure}
  \centering
  \resizebox{0.9\columnwidth}{!}{\includegraphics{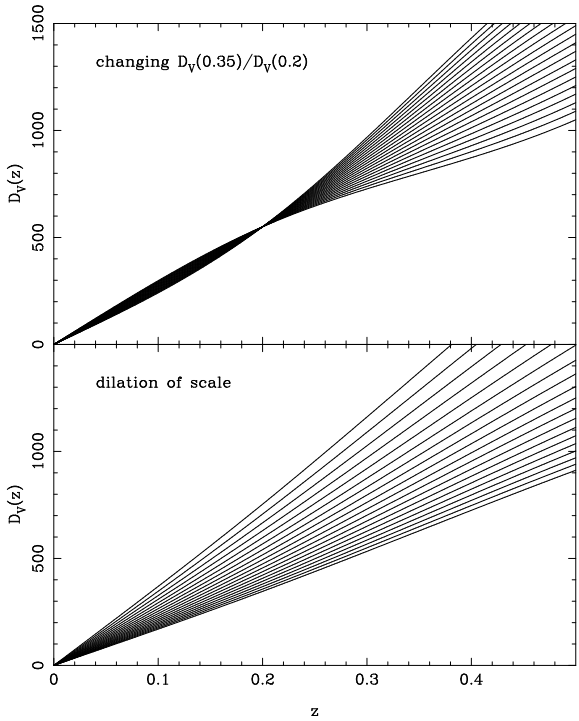}}
  \caption{Two possible ways of changing the distance--redshift model
    tested against the data. Dilating the scale can be achieved by
    simply scaling the measured power spectra and windows, while
    changing the form of the distance--redshift relation requires
    recalculation of the windows. \label{fig:DV_models}}
\end{figure}
We test distance--redshift models that are given by a cubic spline fit
to $D_V$, with one node fixed at $D_V(0.2)=550\mpcoh$ and 41 equally
separated values of another node at $D_V(0.35)$ with
$800<D_V(0.35)<1200\mpcoh$. $D_V(0)=0$ is assumed for each
model. These models are shown in the top panel of
Fig.~\ref{fig:DV_models}. We also allow the distances to be scaled,
which is shown in the lower panel of Fig.~\ref{fig:DV_models} for
fixed $D_V(0.35)/D_V(0.2)$. This scaling can be accomplished without
recalculation of the power spectra, windows or covariances, which can
all be scaled to match the new distance--redshift relation.  In the
spline $\times$ BAO model that we fit to the data, we allow the spline
nodes to vary with this scaling, so that the spline nodes always match
the same locations in the power spectra.

\begin{figure}
  \centering
  \resizebox{0.9\columnwidth}{!}{\includegraphics{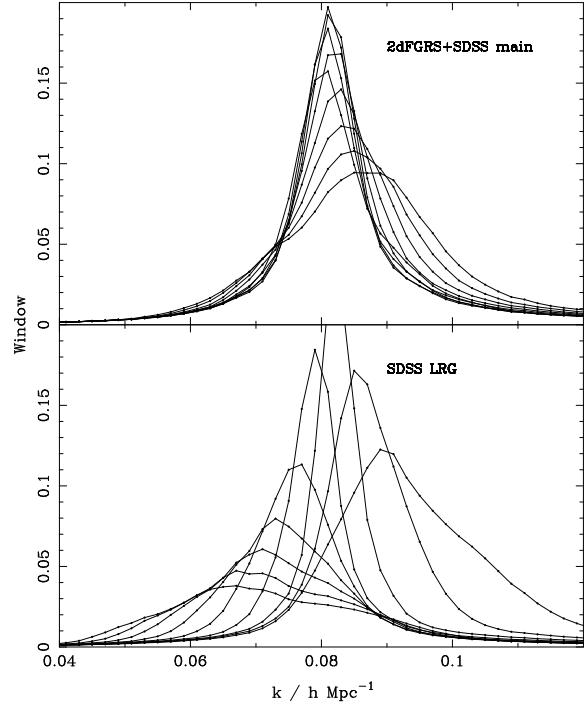}}
  \caption{The window function linking the input power spectrum with
    an observed band-power at $k=0.08\hompc$ (calculated assuming a
    $\Lambda$CDM model), for the SDSS LRG and 2dFGRS + SDSS main
    galaxy catalogues. Window functions are plotted for 9
    distance--redshift models with $D_V(0.2)=550\mpcoh$ and
    $800<D_V(0.35)<1200\mpcoh$. For the LRGs, the peak $k$-value of
    the power that contributes to this measured band--power decreases
    with increasing $D_V(0.35)$. \label{fig:win}}
\end{figure}
Fig.~\ref{fig:bao_lrg_main} shows that we can detect BAO in three
catalogues: SDSS LRG, SDSS main + 2dFGRS and combined SDSS +
2dFGRS. We now provide some of the practical details of how we
constrain the fit to $D_V(z)$ using these data. For each model value
of $D_V(0.35)/D_V(0.2)$, the measured power spectra are a convolution
of the true power, based on the survey geometry and the difference
between the model cosmology and the cosmology used to calculate the
power.  In order to calculate the window function for each
convolution, we have calculated 10000 Gaussian random fields, allowing
the phases and input power spectra to vary. We assume that the true
power is piecewise continuous in bins of width $0.002\hompc$ between
$0<k<0.4\hompc$. We calculated 50 fields where power was only added in
one of these 200 bins. Each field was calculated on a $(512)^3$ grid
covering a $(4000\mpcoh)^3$ cubic volume. Each Gaussian random field
was then translated onto a grid assuming a distance--redshift relation
following the fiducial $\Lambda$CDM cosmology, and is then sampled and
weighted to match the actual survey. The average recovered power
spectrum from each set of 50 realisations then gives part of the
window function of the data given each model, and combining data for
all 200 bins allows the full window function to be
estimated. Fig.~\ref{fig:win} shows a few of the resulting window
functions for the recovered band-power at $k=0.08\hompc$. These models
were calculated with $D_V(0.2)=550\mpcoh$ and 9 values of $D_V(0.35)$
with $800<D_V(0.35)<1200\mpcoh$ with separation $50\mpcoh$. These
numerically determined window functions include both the effects of
the volume covered by the survey, and the different distance--redshift
relation. For the LRGs, when we analyse the data assuming a
$\Lambda$CDM cosmology, if the true value of $D_V(0.35)$ increases,
the scales contributing to a given band-power also increase, and the
peak value in the window function in $k$-space decreases. The
corresponding window functions for the lower redshift data plotted in
the upper panel of Fig.~\ref{fig:win} do not show such a significant
change because the node at $D_V(0.2)$ remains fixed.

We calculate the expected covariances from the LN catalogues described
in Section~\ref{sec:obs_bao}. These catalogues were calculated
allowing for overlap between samples, and power spectra were
calculated as for the actual data. Covariances (internal to each
$P(k)$ and between different power spectra) were recovered assuming
that the power spectra are distributed as a multi-variate
Gaussian. For the set of models tested, we do not change the
covariance matrix with $D_V(0.35)/D_V(0.2)$ (the change in models
shown in the top panel of Fig.~\ref{fig:DV_models}), because the
recovered data power spectra do not change when altering this
parameter combination. Consequently, in this direction, it is the
correlations between data points that primarily change. Tests with
different matrices show that this has a negligible effect across the
set of models, but recalculating the covariance matrices for each
model introduces significant noise into the likelihood surfaces. We do
scale the covariance matrix with the data when we dilate in scale (the
change in models shown in the bottom panel of
Fig.~\ref{fig:DV_models}).

\subsection{Results} \label{sec:results}

In this section, we present likelihood surfaces calculated by fitting
models to the BAO detected in power spectra from the different
samples. In order to remove small likelihood differences caused by
different fits to the overall shape of the power spectrum, we subtract
the likelihood of the best-fit model without BAO from each likelihood
before plotting. The likelihood differences between models with no BAO
is caused by the effect of the different window functions on allowed
shapes of the spline part of the model.

\begin{figure*}
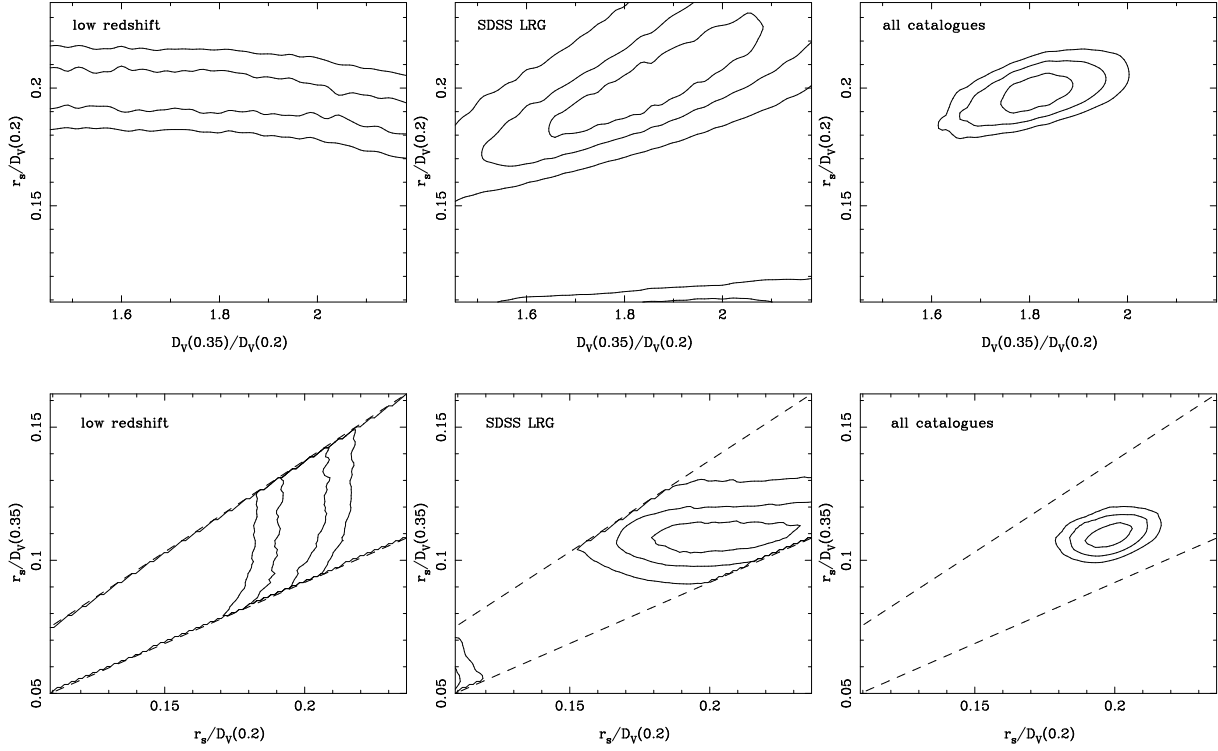

  \centering
  \resizebox{0.3\textwidth}{!}{\includegraphics{like_lowz_dr.ps}}
  \resizebox{0.3\textwidth}{!}{\includegraphics{like_lrg_dr.ps}}
  \resizebox{0.3\textwidth}{!}{\includegraphics{like_all_dr.ps}}
  \\ \vspace{0.5cm}
  \resizebox{0.3\textwidth}{!}{\includegraphics{like_lowz_dd.ps}}
  \resizebox{0.3\textwidth}{!}{\includegraphics{like_lrg_dd.ps}}
  \resizebox{0.3\textwidth}{!}{\includegraphics{like_all_dd.ps}}
  \caption{From left to right: Likelihood surfaces calculated from
    fitting a cubic spline $\times$ BAO model to a single power
    spectrum calculated from the combined main SDSS galaxy + 2dFGRS
    sample, to a single power spectrum calculated using for SDSS LRG
    sample, and to both these power spectra and the additional power
    spectrum calculated from the combined catalogue. Where more than
    one power spectrum is fitted, we allow for correlated errors
    between the power spectra. Likelihood contours were plotted for
    $-2\ln{\cal L}=2.3,\,6.0,\,9.2$, corresponding to two-parameter
    confidence of 68\%, 95\% and 99\% for a Gaussian distribution. In
    the upper row, we plot the contours as a function of
    $r_s/D_V(0.2)$, calculated by dilating the scales of the power
    spectra, windows and covariances, and $D_V(0.35)/D_V(02)$, for
    which different windows were calculated. These likelihoods are
    plotted as a function of $r_s/D_V(0.2)$ and $r_s/D_V(0.35)$ in the
    lower row of this figure. Here the dashed lines show the limits of
    the parameter space tested. \label{fig:like_param1}}
\end{figure*}

Fig.~\ref{fig:like_param1} presents likelihood surfaces calculated by
fitting cubic spline $\times$ BAO models to power spectra calculated
from different sets of data. The upper row of panels show likelihoods
plotted as a function of the 2 parameters used in the analysis,
$D_V(0.35)/D_V(0.2)$, and $r_s/D_V(0.2)$ which is used to parametrise
the dilation of scale. The lower panels show the same likelihood
surfaces after a change of variables to $r_s/D_V(0.2)$ and
$r_s/D_V(0.35)$. BAO within the SDSS main galaxy and 2dFGRS power
spectrum primarily fix the distance to the $z=0.2$, while the LRG
power spectrum measures the distance to $z=0.35$. When we jointly
analyse the power spectra from the low redshift data, the LRGs and the
combination of these samples, we find $r_s/D_V(0.2)=0.1980\pm0.0058$
and $r_s/D_V(0.35)=0.1094\pm0.0033$ (unless stated otherwise all
errors given in this paper are 1-$\sigma$). These constraints are
correlated with correlation coefficient of $0.39$. The likelihood
surface is well approximated by treating these parameters as having a
multi-variate Gaussian distribution with these errors (the left panel
of Fig. \ref{fig:like_test_krange} shows this approximation compared
with the true contours). For completeness, the method for likelihood
calculation is described in Appendix~\ref{app:A}.

For our conventions, $r_s=111.426\mpcoh$ for $\omm=0.25$, $\omb
h^2=0.0223$ and $h=0.72$. Hence, if $\omm h^2 = 0.13$ and $\omb h^2 =
0.0223$, we find $D_V(0.2) = 564\pm23\mpcoh$ and $D_V(0.35) =
1019\pm42\mpcoh$; one can scale to other values of $\omm h^2$ and
$\omb h^2$ using the sound horizon scale formula from Equation
(\ref{eq:rs}).

Without the 2dFGRS data, the low-redshift result reduces to
$r_s/D_V(0.2)=0.1982\pm0.0067$, while the $z=0.35$ result is
unchanged: as expected, the 2dFGRS data primarily help to limit the
distance--redshift relation at $z\sim0.2$.  We can ratio the high and
low redshift BAO position measurements to remove the dependence on the
sound horizon scale $r_S$. From all of the data, we find
$D_V(0.35)/D_V(0.2)=1.812\pm 0.060$. This is higher than the flat
$\Lambda$CDM value, which for $\omm=0.25$ and $\oml=0.75$ is
$D_V(0.35)/D_V(0.2)=1.66$.

\section{testing the method}  \label{sec:tests}

\subsection{The range of scales fitted}

\begin{figure*}
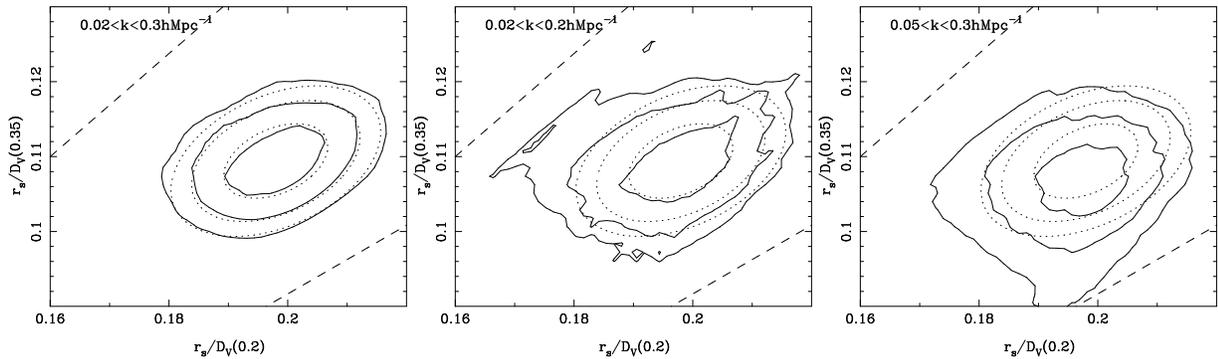

  \centering
  \resizebox{0.3\textwidth}{!}{\includegraphics{like_k0.02.0.3.ps}}
  \resizebox{0.3\textwidth}{!}{\includegraphics{like_k0.02.0.2.ps}}
  \resizebox{0.3\textwidth}{!}{\includegraphics{like_k0.05.0.3.ps}}
  \caption{Likelihood surfaces as plotted in
    Fig.~\ref{fig:like_param1}, but now fitting to different ranges in
    $k$-space (solid contours). As a reference, the dotted contours
    show the Gaussian approximation to the $0.02 < k < 0.3\hompc$
    likelihood surface which has $r_s/D_V(0.2)=0.1980\pm0.0058$ and
    $r_s/D_V(0.35)=0.1094\pm0.0033$, and correlation coefficient of
    $0.39$. Dashed lines show the limit of the parameter ranges
    considered as shown in
    Fig.~\ref{fig:like_param1}. \label{fig:like_test_krange}}
\end{figure*}

Fig.~\ref{fig:like_test_krange} shows the effect of changing the range
of $k$-values fitted on the likelihood surface. Reducing the upper
limit from $k=0.3\hompc$ to $k=0.2\hompc$ does not change the
significance of the best fit, compared to the no-BAO
solution. However, this reduction in the range of $k$ values fitted
increases the possibility of the BAO model fitting spurious noise
because the $0.2<k<0.3\hompc$ data provide a long lever arm to fix the
overall power spectrum shape. Increasing the lower $k$ limit
considered in the fit from $k=0.02\hompc$ to $k=0.05\hompc$ does
reduce the significance of the BAO detection, because the BAO signal
is strongest on large scales. However, there is only a small offset in
the position of the likelihood maximum if we do this, and the
recovered ratio $D_V(0.35)/D_V(0.2)$ is unchanged. This gives us
confidence that we are picking up the oscillatory BAO signal, and that
the large scale features of the BAO, which depend on the details of
the BAO production, do not contribute significantly to the fit.

\subsection{The spline $\times$ BAO model} \label{sec:bao_test}

\begin{figure*}
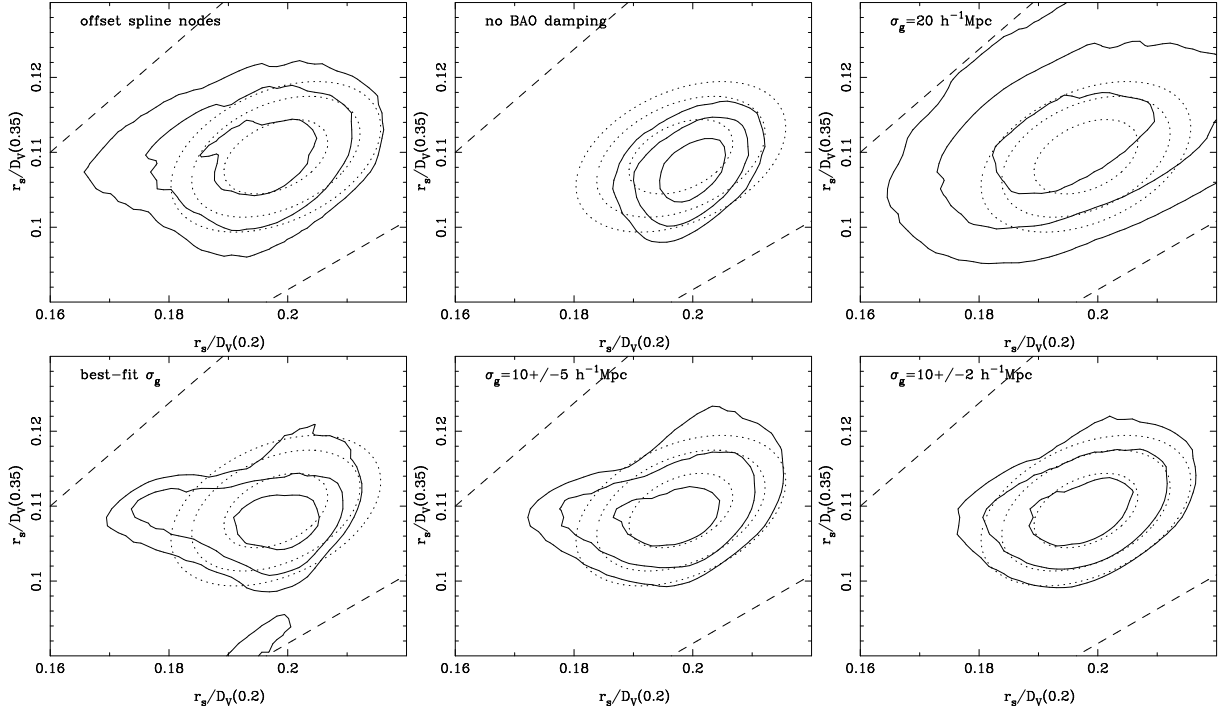

  \centering
  \resizebox{0.3\textwidth}{!}{\includegraphics{like_DV_spline2.ps}}
  \resizebox{0.3\textwidth}{!}{\includegraphics{like_DV_damp00.ps}}
  \resizebox{0.3\textwidth}{!}{\includegraphics{like_DV_damp20.ps}}\\
  \resizebox{0.3\textwidth}{!}{\includegraphics{like_DV_varydamp.ps}}
  \resizebox{0.3\textwidth}{!}{\includegraphics{like_DV_gprior5.ps}}
  \resizebox{0.3\textwidth}{!}{\includegraphics{like_DV_gprior2.ps}}
  \caption{Likelihood surfaces as plotted in
    Fig.~\ref{fig:like_param1}, but now calculated fitting the
    measured power spectra with different spline $\times$ BAO models
    (solid contours). Other lines are as in
    Fig.~\ref{fig:like_test_krange}. Top row, from left to right: we
    consider a spline curve with nodes $k=0.001\hompc$ and
    $k=0.05+n0.05\hompc$ with $n=1,2,...,7$, which are offset in $k$
    compared with our default separation. We use our default spline
    fit, but remove the small-scale BAO damping. We use the default
    spline fit, but increase the position-space BAO damping to be a
    Gaussian with $\sigma_g=20\mpcoh$. Bottom row: likelihood surface
    calculated allowing the damping term, parametrised by $\sigma_g$,
    to float with a uniform prior, and with Gaussian priors
    $\sigma_g=10\pm5\mpcoh$ or
    $\sigma_g=10\pm2\mpcoh$. \label{fig:like_test_spline_model}}
\end{figure*}

Fig. \ref{fig:like_test_spline_model} shows likelihood surfaces
calculated by fitting the measured power spectra with different spline
$\times$ BAO models. We have considered offsetting the nodes of the
spline curve to $k=0.001\hompc$ and 8 nodes with $0.05\le
k\le0.4\hompc$ and separation $\Delta k=0.05\hompc$. Using this form
for the spline curve alters the best-fit parameters to
$r_S/D_V(0.2)=0.1956\pm0.0068$ and
$r_s/D_V(0.35)=0.1092\pm0.0039$. This spline fit is a better match to
the BAO signal on scales $k<0.1\hompc$, leading to a smaller
difference between likelihoods for spline$\times$BAO models and models
with just a spline curve, and larger errors on the recovered
parameters.

\begin{figure}
  \centering
  \resizebox{0.9\columnwidth}{!}{\includegraphics{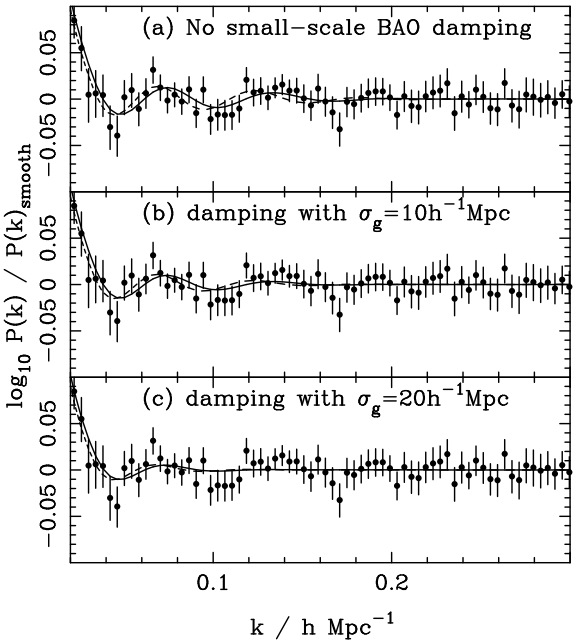}}
  \caption{As Fig.~\ref{fig:bao_lrg_main}, but only for the LRG power
    spectrum, plotted against BAO models with different levels of
    small-scale BAO damping (a) no damping, (b) $\sigma_g=10\mpcoh$,
    (c) $\sigma_g=20\mpcoh$. The solid line is for
    $D_V(0.35)/D_V(0.2)=1.82$, while the dashed line is calculated for
    $D_V(0.35)/D_V(0.2)=1.66$. $D_V(0.2)=568$, matching the values of
    our fiducial $\Lambda$CDM model. \label{fig:bao_damp}}
\end{figure}
Fig. \ref{fig:like_test_spline_model} shows that there is a small
systematic change in the distance ratio $D_V(0.35)/D_V(0.2)$ when the
amplitude of the BAO damping is altered. Increasing the width of the
Gaussian damping model to $\sigma_g=20\mpcoh$ for the BAO fitted to
the three power spectra decreases the best-fit ratio to
$D_V(0.35)/D_V(0.2)=1.769\pm0.079$. Removing the small-scale BAO
damping for all models increases the ratio to
$D_V(0.35)/D_V(0.2)=1.858\pm0.051$. When changing the damping term,
the best fit value of $r_s/D_V(0.2)$ does not change significantly,
and the change in the ratio comes almost entirely from different
fitted values of $r_s/D_V(0.35)$, which is most strongly limited by
the LRG power spectrum. To help to explain this effect,
Fig. \ref{fig:bao_damp} shows BAO models with different values of
$D_V(0.35)/D_V(0.2)$ and damping strength, compared with the observed
LRG BAO. The BAO observed in the LRG power spectrum occur on larger
scales than predicted by our fiducial $\Lambda$CDM model, where
$D_V(0.35)/D_V(0.2)=1.66$. By increasing the strength of the damping,
we reduce the significance of the small-scale signal leading to
increased errors and a ($<1\sigma$) systematic shift to smaller
$D_V(0.35)/D_V(0.2)$. 

If we include $\sigma_g$ as a fitted parameter with a uniform prior,
allowing $\sigma_g$ to vary between power spectra, we obtain best-fit
values $\sigma_g=7.3\pm4.3\mpcoh$ for the low redshift data,
$\sigma_g=1.4\pm2.2\mpcoh$ for the LRGs, and
$\sigma_g=4.7\pm2.6\mpcoh$ for the power spectrum of the combined
sample. Here $D_V(0.35)/D_V(0.2)=1.827\pm0.061$. However, the
inclusion of these extra parameters increases the noise in the
likelihood surfaces. This likelihood surface is shown in
Fig. \ref{fig:like_test_spline_model}, revealing a spur at constant
$r_s/D_V(0.35)$ following models with extreme damping of the low
redshift data, weakening the constraint on $r_s/D_V(0.2)$. The extra
minima at $r_s/D_V(0.35)<0.1$ is due to models with strongly damped
BAO fitted to both the low redshift and combined power
spectra. Likelihood surfaces calculated assuming that $\sigma_g$ has a
Gaussian prior with $\sigma_g=10\pm5\mpcoh$ or $\sigma_g=10\pm2\mpcoh$
are also plotted in Fig. \ref{fig:like_test_spline_model}. As
expected, there is a smooth transition between these likelihood
surfaces, and allowing a small error in $\sigma_g$ does not change the
likelihood significantly from the fixed $\sigma_g=10\mpcoh$ form.

\begin{figure*}
  \centering
  \resizebox{0.3\textwidth}{!}{\includegraphics{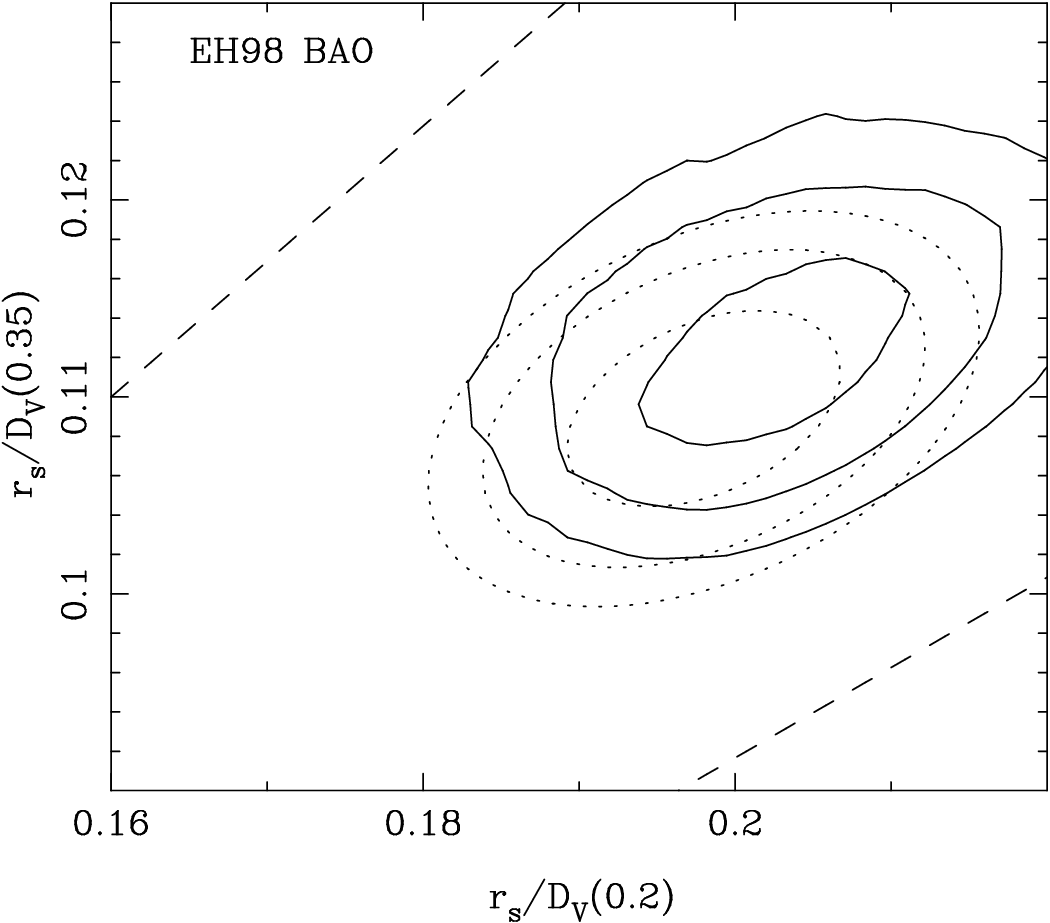}}
  \resizebox{0.3\textwidth}{!}{\includegraphics{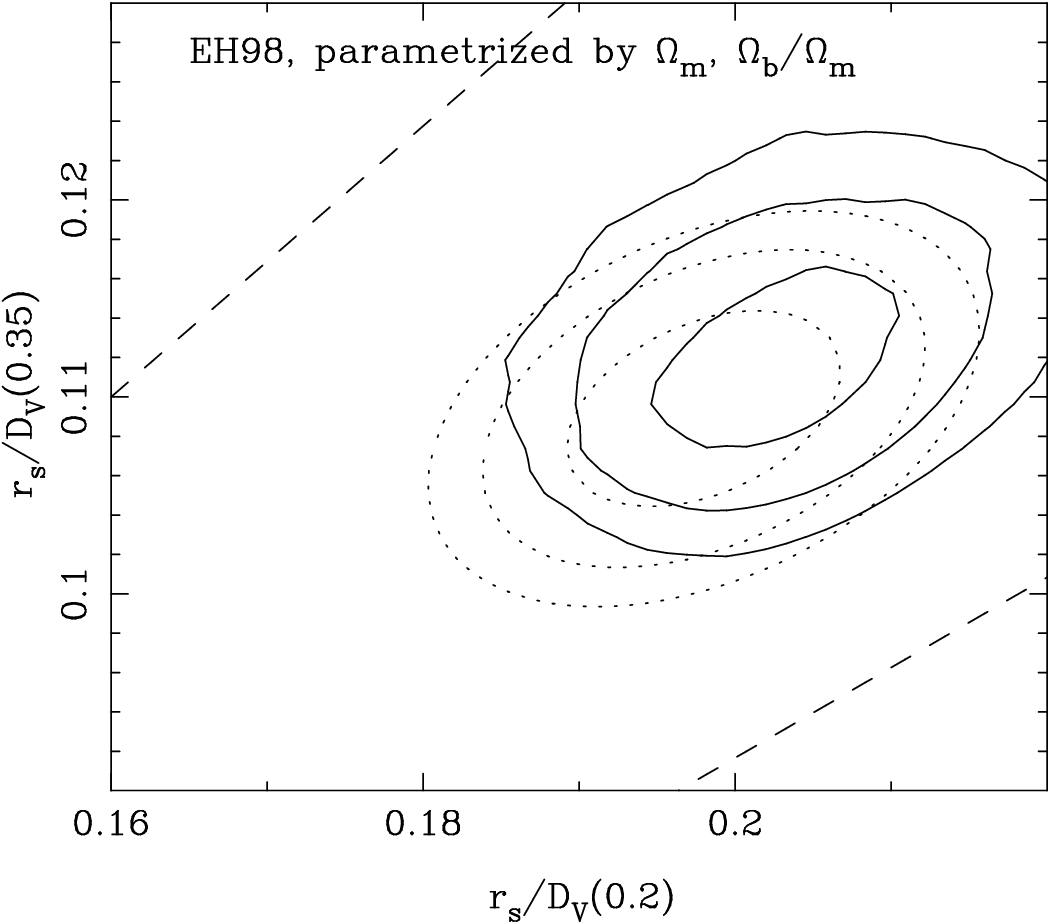}}
  \resizebox{0.3\textwidth}{!}{\includegraphics{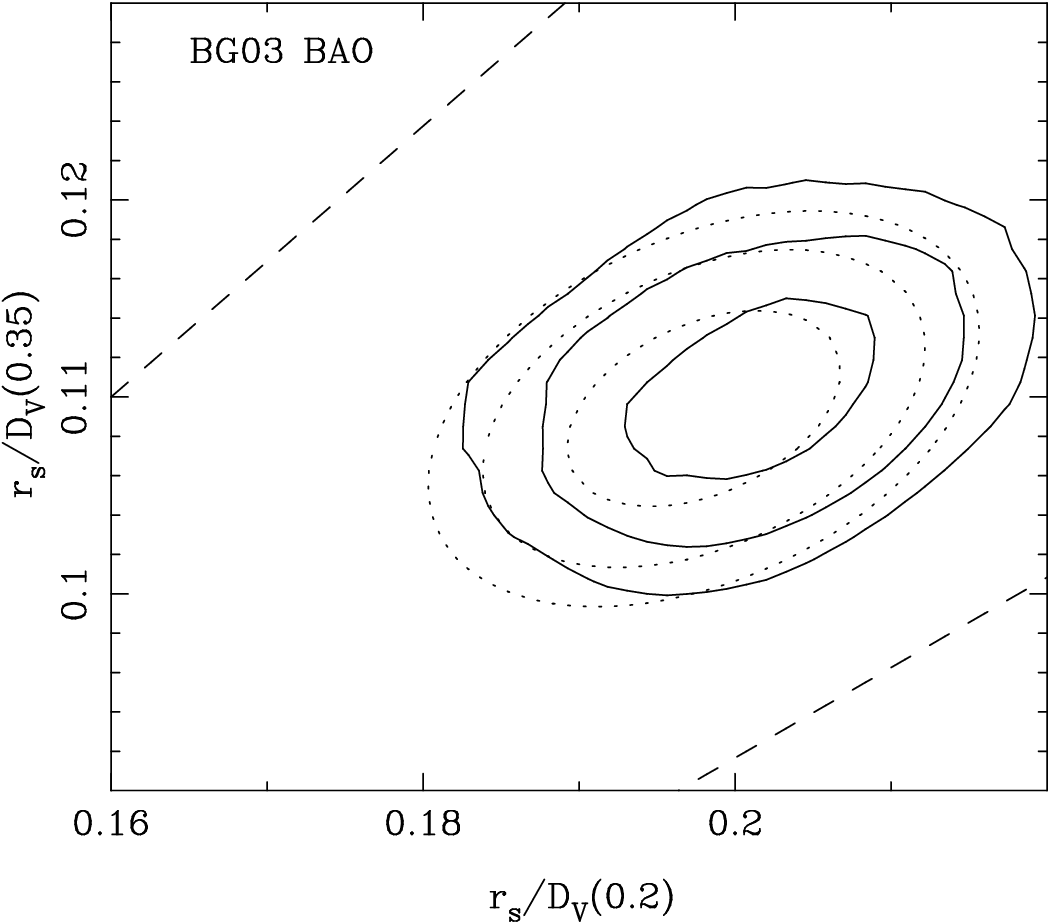}}
  \caption{Likelihood surfaces as plotted in
    Fig.~\ref{fig:like_param1}, but now calculated using different BAO
    models (solid contours). Other lines are as in
    Fig.~\ref{fig:like_test_krange}. From left to right: we use the
    transfer function fits of \citet{eisenstein98} to model the BAO,
    calculated for our fiducial cosmology and stretched in amplitude
    and scale as for the standard {\sc CAMB} model. We again use the
    fits of \citet{eisenstein98} but now allow $\omm$ to change to fix
    the sound horizon scale, and marginalise over the amplitude
    parametrised by $\omb/\omm$. We model the BAO using the simple
    model of Equation
    (\ref{eq:bao_bg03}). \label{fig:like_test_bao_model}}
\end{figure*}

We have also considered how using approximations to the BAO model
affects the fits. Fig.~\ref{fig:like_test_bao_model} shows the
likelihood of different $r_s/D_V(0.2)$ and $r_s/D_V(0.35)$ values,
with BAO models calculated using the \citet{eisenstein98} fitting
formulae, and the simple model of \citet{blake03}, as given by
Equation (\ref{eq:bao_bg03}). The BAO models have been damped assuming
$\sigma_g=10\mpcoh$ for a Gaussian position-space convolution as
described in Section \ref{sec:bao_model}. For the \citet{eisenstein98}
fitting formulae, we have considered two approaches to calculating the
likelihood: either using a fiducial BAO model (calculated for the same
cosmological parameters as our standard {\sc CAMB} model) and
stretching this model in amplitude and scale, or allowing $\omm$ to
vary to match the desired comoving sound horizon scale, and allowing
$\omb/\omm$ to fix the BAO amplitude. The second approach allows the
BAO model on scales $k<0.05\hompc$ to change with cosmological
parameters for fixed value of $D_V(0.2)$. Ideally, in order to
accurately model the BAO on large scales we should separate $r_s$ and
the distance scale in the fits. However, there is little change in the
recovered parameters between these two approaches, demonstrating that
this level of complexity is not required for current data
precision. There is a change in the recovered parameters of order
$<1\sigma$, with best-fit parameters for the \citet{eisenstein98} fits
$r_s/D_V(0.2)=0.2020\pm0.0060$ and $r_s/D_V(0.35)=0.1120\pm0.0033$
with correlation coefficient of $0.41$. For the \citet{blake03} fits,
$r_s/D_V(0.2)=0.2011\pm0.0058$ and $r_s/D_V(0.35)=0.1104\pm0.0034$
with correlation coefficient of $0.37$. The definition of $r_s$ is
built into the \citet{blake03} fit, and will have a different fiducial
value to the other fits.

We might expect the ratio $D_V(0.35)/D_V(0.2)$ to be more robust to
changes in the BAO model as it measures the relative positions of the
BAO at the different redshifts. In essence, by considering this ratio,
we are testing how well the BAO from low and high redshift match. Our
standard {\sc CAMB} fit gave $D_V(0.35)/D_V(0.2)=1.812\pm0.060$. Using
the \cite{eisenstein98} BAO fitting formulae gives
$D_V(0.35)/D_V(0.2)=1.800\pm0.066$, while using the \citet{blake03}
fit gives $D_V(0.35)/D_V(0.2)=1.827\pm0.061$. These are all consistent
at $1\sigma$.

\section{Cosmological Constraints}  \label{sec:cosmo}

\begin{figure}
  \centering
  \resizebox{0.9\columnwidth}{!}{\includegraphics{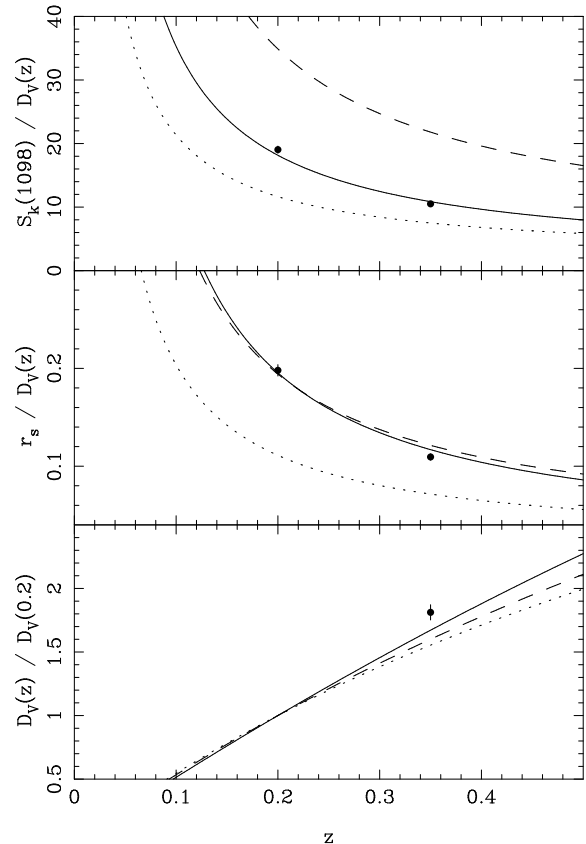}}
  \caption{Three different ways of using BAO to test cosmological
    models. The panels from top to bottom show the constraints on
    $S_k(1098)/D_V(z)$, $r_s/D_V$, and $D_V(z)/D_V(0.2)$ (solid
    circles with $1\sigma$ errors). For many of the data points the
    errors are smaller than the symbols. These data are compared with
    three cosmological models: $\Lambda$CDM ($\omm=0.25$, $\oml=0.75$,
    solid lines), SCDM ($\omm=1$, $\oml=0$, dotted lines), and OCDM
    ($\omm=0.3$, $\oml=0$, dashed lines), as plotted in
    Fig.~\ref{fig:dp_2parfit}. In order to calculate $r_s$ and hence
    $r_s/D_V$, we used the fitting formulae of \citet{eisenstein98},
    assuming $\omb h^2=0.0223$ and $\omm h^2=0.1277$, matching the
    best-fit WMAP numbers for $\Lambda$CDM cosmologies
    \citep{spergel07}. Although the best-fit $D_V(0.35)/D_V(0.2)$
    appears to be further from the $\Lambda$CDM model than in the
    other panels, this is just a consequence of $r_s/D_V(0.2)$ being
    greater than and $r_s/D_V(0.35)$ being less than the $\Lambda$CDM
    model.}
    \label{fig:rsodv}
\end{figure}

We consider three ways of using the BAO scale measurements to restrict
cosmological models. Using just the observed position of the BAO in
the power spectra analysed, we can measure
$D_V(0.35)/D_V(0.2)$. Alternatively, we can compare these distance
scales with the apparent acoustic horizon angle in the CMB: The WMAP
experiment has measured this as $\theta_A=0.5952\pm0.0021^\circ$
\citep{spergel07}. For simplicity, we ignore the $0.4\%$ error on this
measurement, which is negligible compared with the large-scale
structure distance errors, and assume that
$r_s/S_k(1098)=0.0104$. Including this measurement to remove the
dependence on $r_s$ gives $S_k(1098)/D_V(0.2)=19.04\pm0.58$ and
$S_k(1098)/D_V(0.35)=10.52\pm0.32$. The third possibility is that we
model the co-moving sound horizon scale, and simply use the derived
bounds on $r_s/D_V(0.2)$ and $r_s/D_V(0.35)$. This relies on fitting
the comoving sound horizon scale at recombination in addition to the
distance--redshift relation, and has additional parameter dependencies
on $\omm h^2$ and $\omb h^2$. In order to calculate $r_s$ for each
cosmological model tested, we assume that $\omb h^2=0.0223$ and $\omm
h^2=0.1277$, matching the best-fit WMAP numbers for $\Lambda$CDM
cosmologies \citep{spergel07}. We do not include errors on these
parameters, so our recovered errors from fitting $r_s/D_V$ will be
underestimated. The distance ratios $D_V(0.35)/D_V(0.2)$ and
$S_k(1098)/D_V$ are independent of $h$ and $\omb$. These three
possible ways of using the large-scale structure data are shown in
Fig.~\ref{fig:rsodv}, where we compare to three cosmological models.

\begin{figure}
  \centering
  \resizebox{0.9\columnwidth}{!}{\includegraphics{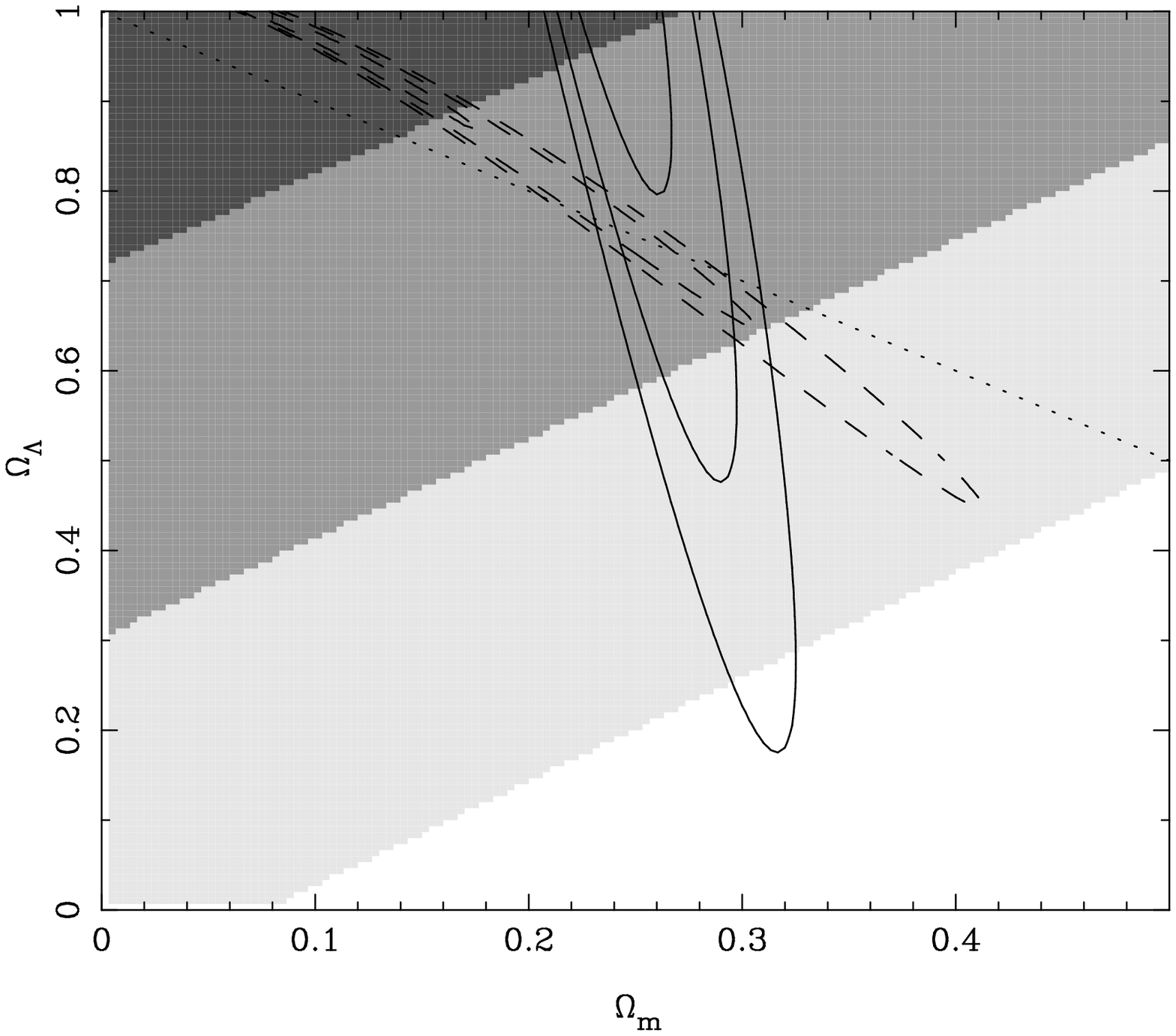}}
  \resizebox{0.9\columnwidth}{!}{\includegraphics{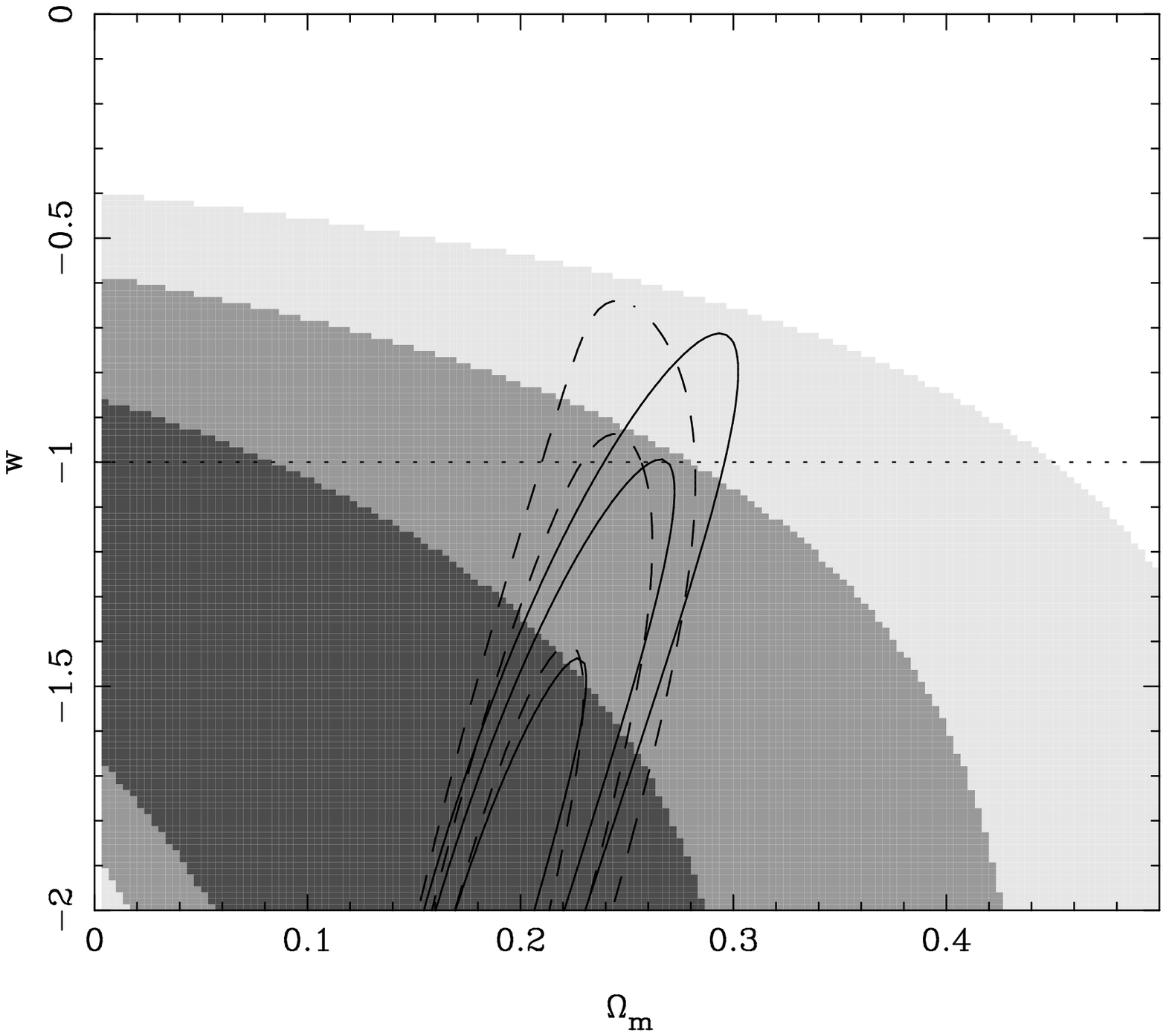}}
  \caption{Top panel: Likelihood surfaces assuming a $\Lambda$CDM
    model parametrised by $\omm$ and $\oml$. Contours and delineations
    between shaded regions are plotted for $-2\ln{\cal
      L}=2.3,\,6.0,\,9.2$. The shaded regions show the likelihood
    given just $D_V(0.35)/D_V(0.2)$. The solid contours were
    calculated by modelling $r_s$ and using constraints on
    $r_s/D_V(0.2)$ and $r_s/D_V(0.35)$, and the dashed contours by
    including the CMB peak position measurement, and use
    $S_k(1098)/D_V(0.2)$ and $S_k(1098)/D_V(0.35)$. The dotted line
    shows the locus of flat models. Bottom panel: likelihood contours
    calculated using the same data, but now for flat cosmological
    models with constant dark energy equation of state parameter
    $w$. Here the dotted line shows $w=-1$. \label{fig:cosmo1}}
\end{figure}

We demonstrate the consistency of the BAO measurements by considering
how they restrict two sets of cosmological models. The top panel of
Fig.~\ref{fig:cosmo1} shows likelihood contours for standard
$\Lambda$CDM cosmologies, parametrised by $\omm$ and $\oml$. The three
ways of using the large-scale structure data that we have considered
constrain different parameter combinations, and the location of their
peak likelihoods do not coincide, although their 95\% confidence
intervals do overlap. In the lower panel we consider flat models with
a constant dark energy equation of state parameter $w$ that is allowed
to vary from $w=-1$. Here, $w<-1$ is favoured at a significance of
$1.4\sigma$, from the $D_V$ ratio assuming a flat prior on $\omm$. 

\begin{figure}
  \centering
  \resizebox{0.9\columnwidth}{!}{\includegraphics{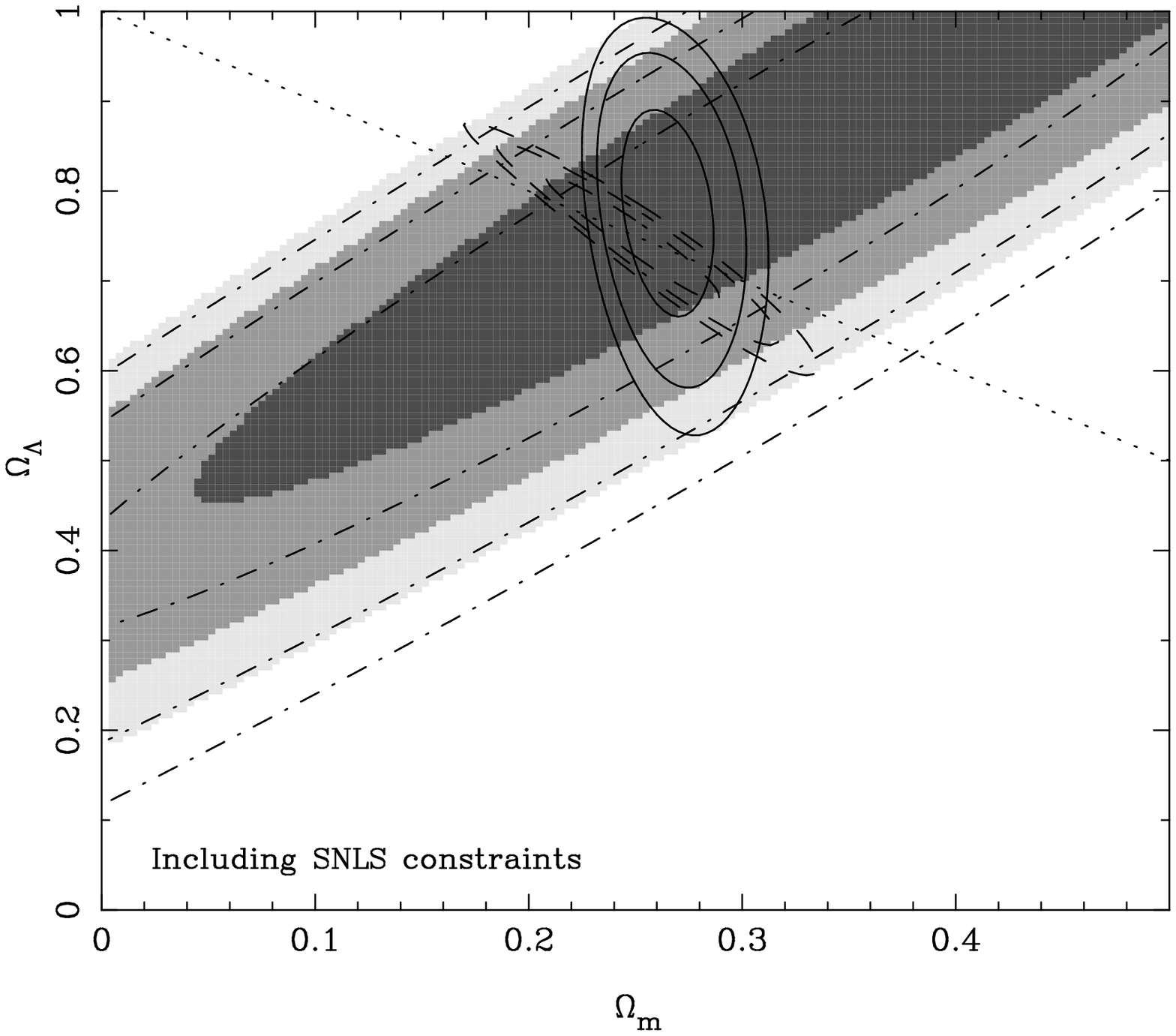}}
  \resizebox{0.9\columnwidth}{!}{\includegraphics{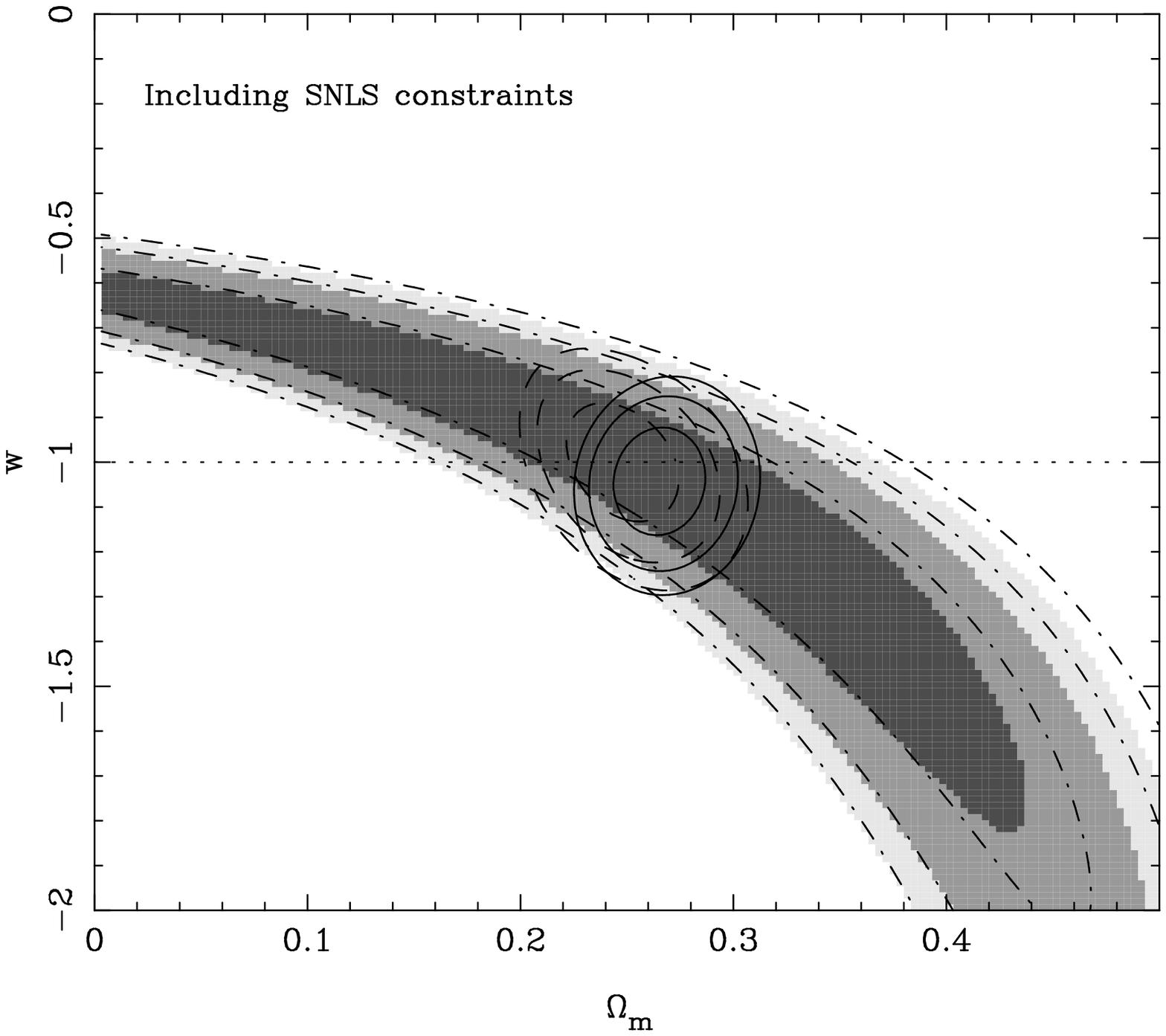}}
  \caption{As Fig.~\ref{fig:cosmo1}, but now additionally using the
    SNIa data presented in \citet{astier06} in the Likelihood
    calculation. The shaded region, dashed and solid contours were
    calculated using the BAO based measurements described in the
    caption to Fig. \ref{fig:cosmo1}. The dot-dashed contours show the
    likelihood surface calculated from just the SNLS
    data. \label{fig:cosmo2}}
\end{figure}

In Fig.~\ref{fig:cosmo2} we have included constraints from the set of
supernovae given in \citet{astier06}. The tightest bounds on models
are obtained if we include the ratio of the sound horizon scale at
recombination to the angular diameter distance to last scattering
calculated from CMB data, which then give a likelihood degeneracy that
is approximately orthogonal to the supernovae likelihood
degeneracy. Including the CMB data gives $\omm=0.252\pm0.027$ and
$\oml=0.743\pm0.047$ for $\Lambda$CDM models. The curvature is found
to be $\omk=-0.004\pm0.022$. For flat models, with constant equation of
state parameter $w$, we find $\omm=0.249\pm0.018$ and
$w=-1.004\pm0.089$.
 
\section{Discussion} \label{sec:discussion}

We have introduced a general method for providing constraints on the
distance--redshift relation using BAO measured from galaxy power
spectra. The method can be applied to different galaxy surveys, or to
subsamples drawn from a single survey that cover different redshift
ranges. At the heart of the method is a likelihood calculation,
matching data and model power spectra, assuming that these have a
multi-variate Gaussian distribution. We now review the components
required for this calculation:
\begin{description}
\item[PARAMETERS] The distance--redshift relation is parametrized
  using a spline fit in $D_V(z)$ with a small number of nodes,
  $D_V(z_i)$. We can simply scale measured power spectra to follow a
  multiplicative shift of all $D_V(z_i)$, so we take as parameters
  $D_V(z_i)/D_V(z_j)$, $i\ne j$, and $D_V(z_1)$ (this was discussed in
  Section~\ref{sec:obs_bao_scale}). This parametrization allows the
  results to be used to constrain general cosmological models that
  have such a smooth $D_V(z)$, without having to specify the set of
  models before the start of the analysis.
\item[DATA] The galaxies are split into subsamples covering different
  (possibly overlapping) redshift ranges. The power spectra for these
  samples are calculated assuming a fiducial cosmological model
  (Section~\ref{sec:obs_bao_scale}). The position of the BAO in each
  power spectrum depends on a weighted integral of the
  distance--redshift relation for the range of redshifts covered by
  the sample from which the power spectrum is
  calculated. Consequently, by fitting power spectra from different
  samples, we can measure the ratio of distances to different
  redshifts.
\item[BAO MODEL] BAO are extracted from a model power spectrum
  calculated using {\sc CAMB}, by fitting with a spline $\times$ BAO
  model, as fitted to the observed galaxy power spectra. These BAO are
  stretched to allow for varying $r_s/D_V(z_1)$
  (Section~\ref{sec:obs_bao}).
\item[MODEL] The model is formed from a smooth spline curve multiplied
  by the BAO model (Section~\ref{sec:obs_bao}). This is convolved with
  the window function, which corrects for both the survey geometry,
  and the difference between the fiducial cosmology (at which the data
  power spectra were calculated), and the cosmological model to be
  tested (Section~\ref{sec:basic_method}). The window functions were
  calculated using realisations of Gaussian random fields.
\item[ERRORS] Covariance matrices for the power spectra were
  calculated from Log-Normal realisations of galaxy
  distributions. Covariances between the different power spectra of
  different galaxy samples were included.
\item[NUISANCE PARAMETERS] The spline nodes giving the shapes of the
  power spectra were fixed at their best fit values for each model
  tested. We are therefore left with a likelihood measurements for a
  set of $r_s/D_V(z_1)$ and $D_V(z_i)/D_V(z_j)$, $i\ne j$ values.
\end{description}

This analysis method has been used to jointly analyse samples of
galaxies drawn from the SDSS and 2dFGRS. BAO were calculated by
fitting a fiducial power spectrum calculated by {\sc CAMB}
\citep{lewis00}. We have considered using fitting formulae to
calculate the BAO \citep{eisenstein98,blake03}, and find changes in
the recovered BAO scale of order $1\sigma$. Such a dependence was also
found recently by \citealt{angulo07} when fitting simulated data, and
it is clear that the combined 2dFGRS+SDSS data now reveal the BAO with
sufficient accuracy that we need to take care when modelling the BAO.

The BAO scale measurements were used to set limits on two sets of
cosmological models: Standard $\Lambda$ models, and flat models with
constant dark energy equation of state. When we analyse flat
$\Lambda$CDM models, we find similar errors on the matter density to
those obtained by \citet{percival07a}, where these models were
directly compared with the data. The SNIa data from \citet{astier06}
provide cosmological constraints that have a similar degeneracy
direction to the lower redshift BAO constraint on
$D_V(0.35)/D_V(0.2)$. However, if we include the information from the
position of the peak in the WMAP CMB data, or model the sound horizon
scale at recombination then the likelihoods become
complementary. These two approaches provide different best-fit
parameters, although they are consistent at the 1$\sigma$ level. For
$\Lambda$CDM models $\omm=0.266\pm0.015$ if we model the sound horizon
scale, or $\omm=0.252\pm0.027$ including the CMB data. Similarly, for
flat models with constant $w$, we find $w=-1.045\pm0.080$ if we model
the sound horizon scale, or $w=-1.004\pm0.088$ including the CMB data.

For flat models with constant $w$, the differential distance
measurement $D_V(0.35)/D_V(0.2)$ favours $w<-1$. However, it is worth
noting that Fig.~\ref{fig:cosmo1} shows that the total density
($\Omega_{\rm tot}$) and $w$ are highly coupled, so allowing curvature
to vary would significantly weaken this conclusion \citep{clarkson07}.
The SNLS supernovae data favour $w\simeq-1$, hinting at a discrepancy
between low and high redshift. Fitting to the SNLS SNIa data gives
$D_V(0.35)/D_V(0.2)=1.666\pm0.010$ for the set of $\Lambda$CDM models
considered, or $D_V(0.35)/D_V(0.2)=1.665\pm0.010$ for flat models with
constant dark energy equation of state.

The tests presented in Section~\ref{sec:bao_test} show that the
measured distance ratio from the current BAO data is sensitive to the
damping model. This is clear from Fig. \ref{fig:bao_damp}, where it is
apparent that there is a small offset between all models and the
positions of the first and second peaks in the LRG BAO.  By increasing
the BAO damping, we decrease the significance of the second peak
compared with the first, and change the fitted ratio
$D_V(0.35)/D_V(0.2)$. However, our default choice of the damping model
-- a Gaussian convolution in position space with
$\sigma_g\sim10\mpcoh$ -- is well motivated by current simulation
results \citep{eisenstein07,angulo07}. This gives
$D_V(0.35)/D_V(0.2)=1.812\pm 0.060$, which is offset by $2.4\sigma$
from the SNIa results. If this is not a case of extreme bad luck, we
must therefore consider at least one of the following options:
\begin{enumerate}
\item The damping model needs to be revised and made more
sophisticated;
\item The data/analysis is flawed in a way that evades the tests
we have performed so far;
\item The simple $\Lambda$ model is wrong.
\end{enumerate}

For the Gold supernovae data set \citep{riess04}, the significance of
any evidence for $w<-1$ at low redshift would increase because this
SNIa dataset also favours strong dark energy at $z<0.3$ -- so
it is conceivable that
this discrepancy could be
genuinely cosmological in origin. However, in this paper we
only compare with the SNLS data because of the benefits of considering
homogeneous data. It will be interesting to recalculate this
significance when the SDSS supernova survey \citep{nichol07} is
complete, as it focuses on $z<0.5$, and should either
confirm or reject any deviations from a simple $\Lambda$CDM model at
these low redshifts.

\section*{Acknowledgements}

WJP is grateful for support from a PPARC advanced fellowship. WJP
acknowledges useful conversations with Sanjeev Seahra, and
constructive comments from David Weinberg on an early draft of this
manuscript. Simulated catalogues were calculated and analysed using
the COSMOS Altix 3700 supercomputer, a UK-CCC facility supported by
HEFCE and PPARC in cooperation with CGI/Intel.

The 2dF Galaxy Redshift Survey was undertaken using the Two-degree Field
facility on the 3.9m Anglo-Australian Telescope. The success of the survey
was made possible by the dedicated efforts of the staff of the
Anglo-Australian Observatory, both in creating the 2dF instrument and
in supporting the survey observations.

Funding for the SDSS and SDSS-II has been provided by the Alfred
P. Sloan Foundation, the Participating Institutions, the National
Science Foundation, the U.S. Department of Energy, the National
Aeronautics and Space Administration, the Japanese Monbukagakusho, the
Max Planck Society, and the Higher Education Funding Council for
England. The SDSS Web Site is {\tt http://www.sdss.org/}.

The SDSS is managed by the Astrophysical Research Consortium for the
Participating Institutions. The Participating Institutions are the
American Museum of Natural History, Astrophysical Institute Potsdam,
University of Basel, Cambridge University, Case Western Reserve
University, University of Chicago, Drexel University, Fermilab, the
Institute for Advanced Study, the Japan Participation Group, Johns
Hopkins University, the Joint Institute for Nuclear Astrophysics, the
Kavli Institute for Particle Astrophysics and Cosmology, the Korean
Scientist Group, the Chinese Academy of Sciences (LAMOST), Los Alamos
National Laboratory, the Max-Planck-Institute for Astronomy (MPIA),
the Max-Planck-Institute for Astrophysics (MPA), New Mexico State
University, Ohio State University, University of Pittsburgh,
University of Portsmouth, Princeton University, the United States
Naval Observatory, and the University of Washington.

\setlength{\bibhang}{2.0em}

\appendix

\section{Likelihood Calculation}  \label{app:A}

The best fit parameters from our analysis of BAO are
$r_s/D_V(0.2)=0.1980\pm0.0058$ and $r_s/D_V(0.35)=0.1094\pm0.0033$,
with correlation coefficient of $0.39$. A multi-variate Gaussian
likelihood can be estimated from using these numbers given model
values of $r_s/D_V(0.2)$ and $r_s/D_V(0.35)$ as $-2\ln{\cal L} \propto
\bf{X}^{-1}{\bf V}^{-1}\bf{X}$, where
\begin{eqnarray}
  \bf{X} &=& \left(\begin{array}{c} 
    \frac{r_s}{D_V(0.2)}  - 0.1980 \\
    \frac{r_s}{D_V(0.35)} - 0.1094 \end{array}
  \right), \\
  {\bf V}^{-1} &=& \left(\begin{array}{cc} 
     35059 & -24031 \\
    -24031 & 108300 \end{array}
  \right).
\end{eqnarray}

\label{lastpage}

\end{document}